\documentclass[twocolumn]{aastex631}

\shorttitle{\textit{JWST} Observations of VHS 1256 b}
\shortauthors{Miles et al.}
\usepackage{graphicx}
\graphicspath{{./}{figures/}}
\usepackage{lineno}

\begin{document}

\title{The JWST Early Release Science Program for Direct Observations of Exoplanetary Systems II: A 1 to 20 Micron Spectrum of the Planetary-Mass Companion VHS 1256-1257 b}

\author[0000-0002-5500-4602]{Brittany E.~Miles   }\affiliation{Department of Astronomy and Astrophysics, University of California, Santa Cruz, 1156 High St, Santa Cruz, CA 95064}\affiliation{Department of Physics and Astronomy, University of California, Irvine, 4129 Frederick Reines Hall, Irvine, CA 92697-4575}

\author[0000-0003-4614-7035]{Beth A.~Biller      }\affiliation{Scottish Universities Physics Alliance, Institute for Astronomy, University of Edinburgh, Blackford Hill, Edinburgh EH9 3HJ, UK; Centre for Exoplanet Science, University of Edinburgh, Edinburgh EH9 3HJ, UK}

\author[0000-0001-8718-3732]{Polychronis Patapis }\affiliation{Institute of Particle Physics and Astrophysics, ETH Zurich, Wolfgang-Pauli-Str. 27, 8093 Zurich, Switzerland}
\author[0000-0002-5885-5779]{Kadin Worthen	  }\affiliation{Department of Physics and Astronomy, John's Hopkins University, 3400 N. Charles Street, Baltimore, MD 21218, USA}	
\author[0000-0003-4203-9715]{Emily Rickman       }\affiliation{European Space Agency (ESA), ESA Office, Space Telescope Science Institute, 3700 San Martin Drive, MD 21218, USA}
\author[0000-0002-9803-8255]{Kielan K.~W.~Hoch	 }\affiliation{Space Telescope Science Institute, Baltimore, MD 21218, USA}\affiliation{Center for Astrophysics and Space Sciences,  University of California, San Diego, La Jolla, CA 92093, USA}
\author[0000-0001-6098-3924]{Andrew Skemer       }\affiliation{Department of Astronomy and Astrophysics, University of California, Santa Cruz, 1156 High St, Santa Cruz, CA 95064}
\author[0000-0002-3191-8151]{Marshall D.~Perrin  }\affiliation{Space Telescope Science Institute, Baltimore, MD 21218, USA}
\author[0000-0001-8818-1544]{Niall Whiteford     }\affiliation{Department of Astrophysics, American Museum of Natural History, Central Park West at 79th St., New York, NY 10024, USA}
\author[0000-0002-8382-0447]{Christine H.~Chen	  }\affiliation{Space Telescope Science Institute, 3700 San Martin Dr., Baltimore, MD 21218, USA}\affiliation{Department of Physics and Astronomy, John's Hopkins University, 3400 N. Charles Street, Baltimore, MD 21218, USA}
\author[0000-0001-9855-8261]{B.~Sargent	         }\affiliation{Space Telescope Science Institute, 3700 San Martin Dr., Baltimore, MD 21218, USA}

\author[0000-0003-1622-1302]{Sagnick Mukherjee	 }\affiliation{Department of Astronomy and Astrophysics, University of California, Santa Cruz, 1156 High St, Santa Cruz, CA 95064}
\author[0000-0002-4404-0456]{Caroline V.~Morley	 }\affiliation{Department of Astronomy, The University of Texas at Austin, Austin, TX 78712, USA}	
\author[0000-0002-6721-3284]{Sarah E.~Moran	  }\affiliation{Dept.\ of Planetary Sciences; Lunar \& Planetary Laboratory; Univ.\ of Arizona; Tucson, AZ 85721}

\author[0000-0001-5579-5339]{Mickael Bonnefoy    }\affiliation{Universit\'{e} Grenoble Alpes, Institut de Plan\'{e}tologie et d'Astrophysique (IPAG), F-38000 Grenoble, France}	
\author[0000-0003-0331-3654]{Simon Petrus	 }\affiliation{Instituto de F\'{i}sica  y Astronom\'{i}a Facultad de Ciencias, Universidad de Valpara\'{i}so,  Av. Gran Breta\~{n}a 1111, Valpara\'{i}so, Av. Gran Breta\~{n}a 1111, Valpara\'{i}so, Chile} \affiliation{N\'{u}cleo Milenio Formac\'{i}on Planetaria - NPF, Universidad de Valpara\'{i}so, Av. Gran Breta\~{n}a 1111, Valpara\'{i}so, Chile}
\author[0000-0001-5365-4815]{Aarynn L.~Carter    }\affiliation{Department of Astronomy and Astrophysics, University of California, Santa Cruz, 1156 High St, Santa Cruz, CA 95064}
\author[0000-0002-9173-0740]{Elodie Choquet      }\affiliation{Aix Marseille Univ, CNRS, CNES, LAM, Marseille, France}	
\author[0000-0001-8074-2562]{Sasha Hinkley	 }\affiliation{University of Exeter, Astrophysics Group, Physics Building, Stocker Road, Exeter, EX4 4QL, UK}

\author[0000-0002-4479-8291]{Kimberly Ward-Duong }\affiliation{Department of Astronomy, Smith College, Northampton, MA, 01063, USA}	
\author[0000-0002-0834-6140]{Jarron M.~Leisenring}\affiliation{Steward Observatory, University of Arizona, 933 N. Cherry Ave., Tucson, AZ 85721, USA}	
\author[0000-0001-6205-9233]{Maxwell A.~Millar-Blanchaer}\affiliation{Department of Physics, University of California, Santa Barbara, CA, 93106}

\author{Laurent Pueyo        }\affiliation{Space Telescope Science Institute, Baltimore, MD 21218, USA}	
\author[0000-0003-2259-3911]{Shrishmoy Ray       }\affiliation{University of Exeter, Astrophysics Group, Physics Building, Stocker Road, Exeter, EX4 4QL, UK}	
\author{Steph Sallum        }\affiliation{ Department of Physics and Astronomy, University of California, Irvine, 4129 Frederick Reines Hall, Irvine, CA 92697-4575}
\author[0000-0002-2805-7338]{Karl R.~Stapelfeldt }\affiliation{Jet Propulsion Laboratory, California Institute of Technology, Mail Stop 321-100, 4800 Oak Grove Drive, Pasadena CA 91109  USA}	
\author[0000-0003-0454-3718]{Jordan M.~Stone     }\affiliation{Naval Research Laboratory, Remote Sensing Division, 4555 Overlook Ave SW, Washington, DC 20375 USA}	
\author[0000-0003-0774-6502]{Jason J.~Wang       }\affiliation{Center for Interdisciplinary Exploration and Research in Astrophysics (CIERA) and Department of Physics and Astronomy, Northwestern University, Evanston, IL 60208, USA} \affiliation{Department of Astronomy, California Institute of Technology, Pasadena, CA 91125, USA}

\author[0000-0002-4006-6237]{Olivier Absil	  }\affiliation{STAR Institute, Universit\'e de Li\`ege, All\'ee du Six Ao\^{u}t 19c, 4000 Li\`ege, Belgium}
\author[0000-0001-6396-8439]{William O.~Balmer    }\affiliation{Department of Physics and Astronomy, John's Hopkins University, 3400 N. Charles Street, Baltimore, MD 21218, USA}\affiliation{Space Telescope Science Institute, Baltimore, MD 21218, USA}		
\author[0000-0001-9353-2724]{Anthony Boccaletti	  }\affiliation{LESIA, Observatoire de Paris, Universit{\'e} PSL, CNRS, Universit{\'e} Paris Cit{\'e}, Sorbonne Universit{\'e}, 5 plaƒce Jules Janssen, 92195 Meudon, France}	
\author[0000-0002-7520-8389]{Mariangela Bonavita  }\affiliation{School of Physical Sciences, Faculty of Science, Technology, Engineering and Mathematics, The Open University, Walton Hall, Milton Keynes, MK7 6AA}	
\author[0000-0001-8568-6336]{Mark Booth	          }
\affiliation{Astrophysikalisches Institut und Universit\"atssternwarte, Friedrich-Schiller-Universit\"at Jena, Schillerg\"a\ss{}chen}
2-3, D-07745 Jena, Germany
\author[0000-0003-2649-2288]{Brendan P.~Bowler	  }\affiliation{Department of Astronomy, The University of Texas at Austin, Austin, TX 78712, USA}	
\author[0000-0003-4022-8598]{Gael Chauvin	  }\affiliation{Laboratoire J.-L. Lagrange, Universit\'e Cote dÕAzur, CNRS, Observatoire de la Cote dÕAzur, 06304 Nice, France}
\author[0000-0002-0101-8814]{Valentin Christiaens }\affiliation{STAR Institute, Universit\'e de Li\`ege, All\'ee du Six Ao\^{u}t 19c, 4000 Li\`ege, Belgium}
\author[0000-0002-7405-3119]{Thayne Currie	  }\affiliation{Department of Physics and Astronomy, University of Texas-San Antonio, 1 UTSA Circle, San Antonio, TX, USA; Subaru Telescope, National Astronomical Observatory of Japan,  650 North A`oh$\bar{o}$k$\bar{u}$ Place, Hilo, HI  96720, USA}	
\author[0000-0002-3729-2663]{Camilla Danielski	  }\affiliation{Instituto de Astrof\'isica de Andaluc\'ia, CSIC, Glorieta de la Astronom\'ia s/n, 18008, Granada, Spain}	
\author[0000-0002-9843-4354]{Jonathan J.~Fortney  }\affiliation{Department of Astronomy and Astrophysics, University of California, Santa Cruz, 1156 High St, Santa Cruz, CA 95064}	
\author[0000-0001-8627-0404]{Julien H.~Girard     }\affiliation{Space Telescope Science Institute, Baltimore, MD 21218, USA}	
\author{Carol A.~Grady	  }\affiliation{Eureka Scientific, 2452 Delmer. St., Suite 1, Oakland CA, 96402, United States}	
\author[0000-0002-7162-8036]{Alexandra Z.~Greenbaum}\affiliation{IPAC, California Institute of Technology, 1200 E. California Boulevard, Pasadena, CA 91125, USA}	
\author[0000-0002-1493-300X]{Thomas Henning	  }\affiliation{MPI for Astronomy, K\""onigstuhl 17, 69117 Heidelberg, Germany}	
\author[0000-0003-4653-6161]{Dean C.~Hines	  }\affiliation{Space Telescope Science Institute, Baltimore, MD 21218, USA}	
\author[0000-0001-8345-593X]{Markus Janson	  }\affiliation{Department of Astronomy, Stockholm University, AlbaNova University Center, SE-10691 Stockholm}

\author[0000-0002-6221-5360]{Paul Kalas	          }\affiliation{Department of Astronomy, 501 Campbell Hall, University of California Berkeley, Berkeley, CA 94720-3411, USA} \affiliation{SETI Institute, Carl Sagan Center, 189 Bernardo Ave.,  Mountain View CA 94043, USA} \affiliation{Institute of Astrophysics, FORTH, GR-71110 Heraklion, Greece}	
\author[0000-0003-2769-0438]{Jens Kammerer	  }\affiliation{Space Telescope Science Institute, Baltimore, MD 21218, USA}	\author[0000-0001-6831-7547]{Grant M.~Kennedy	  }\affiliation{Department of Physics, University of Warwick, Gibbet Hill Road, Coventry, CV4 7AL, UK}	
\author[0000-0002-7064-8270]{Matthew A.~Kenworthy }\affiliation{Leiden Observatory, Leiden University, P.O. Box 9513, 2300 RA Leiden, The Netherlands}	
\author[0000-0003-0626-1749]{Pierre Kervella	  }\affiliation{LESIA, Observatoire de Paris, Universit{\'e} PSL, CNRS, Universit{\'e} Paris Cit{\'e}, Sorbonne Universit{\'e}, 5 place Jules Janssen, 92195 Meudon, France}	
\author{Pierre-Olivier Lagage}\affiliation{Universit{\'e} Paris-Saclay, Universit{\'e} Paris Cit{\'e}, CEA, CNRS, AIM, 91191, Gif-sur-Yvette, France}	
\author[0000-0003-1487-6452]{Ben W.~P.~Lew	        }\affiliation{Bay Area Environmental Research Institute and NASA Ames Research Center, Moffett Field, CA 94035, USA}
\author[0000-0003-2232-7664]{Michael C.~Liu	  }\affiliation{Institute for Astronomy, University of Hawai'i, 2680 Woodlawn Drive, Honolulu HI 96822}	
\author[0000-0003-1212-7538]{Bruce Macintosh	  }\affiliation{Kavli Institute for Particle Astrophysics and Cosmology, Stanford University, Stanford California 94305}	\affiliation{University of California Observatories}
\author[0000-0002-5352-2924]{Sebastian Marino	  }\affiliation{Jesus College, University of Cambridge, Jesus Lane, Cambridge CB5 8BL, UK}\affiliation{Institute of Astronomy, University of Cambridge, Madingley Road, Cambridge CB3 0HA, UK}\affiliation{University of Exeter, Astrophysics Group, Physics Building, Stocker Road, Exeter, EX4 4QL, UK}
\author[0000-0002-5251-2943]{Mark S.~Marley	  }\affiliation{Dept.\ of Planetary Sciences; Lunar \& Planetary Laboratory; Univ.\ of Arizona; Tucson, AZ 85721}	
\author[0000-0002-4164-4182]{Christian Marois	  }\affiliation{National Research Council of Canada}	
\author[0000-0003-0593-1560]{Elisabeth C.~Matthews}\affiliation{Max-Planck-Institut f\"ur Astronomie, K\"onigstuhl 17, 69117 Heidelberg, Germany}	\affiliation{Observatoire de l'Universit\'e de Gen\`eve, Chemin Pegasi 51, 1290 Versoix, Switzerland}	
\author[0000-0003-3017-9577]{Brenda  C.~Matthews  }\affiliation{Herzberg Astronomy \& Astrophysics Research Centre, National Research Council of Canada, 5071 West Saanich Road, Victoria, BC V9E 2E7, Canada}	
\author[0000-0002-8895-4735]{Dimitri Mawet	  }\affiliation{Department of Astronomy, California Institute of Technology, Pasadena, CA 91125, USA}\affiliation{Jet Propulsion Laboratory, California Institute of Technology, 4800 Oak Grove Dr.,Pasadena, CA 91109, USA}	
\author[0000-0003-0241-8956]{Michael W.~McElwain  }\affiliation{Exoplanets and Stellar Astrophysics Laboratory, NASA Goddard Space Flight Center, Greenbelt, MD 20771, USA}	
\author[0000-0003-3050-8203]{Stanimir Metchev	  }\affiliation{Western University, Department of Physics \& Astronomy and Institute for Earth and Space Exploration, 1151 Richmond Street, London, Ontario N6A 3K7, Canada}	
\author[0000-0003-1227-3084]{Michael R.~Meyer	  }\affiliation{Department of Astronomy, University of Michigan, 1085 S. University, Ann Arbor, MI 48109, USA}	
\author[0000-0003-4096-7067]{Paul Molliere	  }\affiliation{Max-Planck-Institut f\"ur Astronomie, K\"onigstuhl 17, 69117 Heidelberg, Germany}	
\author[0000-0001-6472-2844]{Eric Pantin          }\affiliation{IRFU/DAp D\'epartement D'Astrophysique CE Saclay, Gif-sur-Yvette, France}
\author{Andreas Quirrenbach  }\affiliation{Landessternwarte, Zentrum f\"ur Astronomie der Universit\"at Heidelberg, K\"onigstuhl 12, D-69117 Heidelberg, Germany}	
\author[0000-0002-4388-6417]{Isabel Rebollido	  }\affiliation{Space Telescope Science Institute, Baltimore, MD 21218, USA}	
\affiliation{Centro de Astrobiolog\'ia (CAB, CSIC-INTA) , ESAC Campus Camino Bajo del Castillo, s/n, Villanueva de la Ca\~nada, E-28692 Madrid, Spain
}

\author[0000-0003-1698-9696]{Bin B.~Ren	          }\affiliation{Universit\'{e} Grenoble Alpes, Institut de Plan\'{e}tologie et d'Astrophysique (IPAG), F-38000 Grenoble, France}	
\author{Glenn Schneider      }\affiliation{Steward Observatory, University of Arizona, 933 N. Cherry Ave., Tucson, AZ 85721, USA}
\author[0000-0002-4511-3602]{Malavika Vasist	  }\affiliation{STAR Institute, Universit\'e de Li\`ege, All\'ee du Six Ao\^{u}t 19c, 4000 Li\`ege, Belgium}

\author[0000-0001-9064-5598]{Mark C.~Wyatt	  }\affiliation{Institute of Astronomy, University of Cambridge, Madingley Road, Cambridge, CB3 0HA, UK}	
\author[0000-0003-2969-6040]{Yifan Zhou	          }\affiliation{Department of Astronomy, The University of Texas at Austin, Austin, TX 78712, USA}

\author[0000-0002-1764-2494]{Zackery W. Briesemeister}\affiliation{NASA Goddard Space Flight Center, Greenbelt, MD 20771, USA}	
\author[0000-0002-6076-5967]{Marta L Bryan	           }\affiliation{Department of Astronomy, 501 Campbell Hall, University of California Berkeley, Berkeley, CA 94720-3411, USA}	
\author[0000-0002-5335-0616]{Per Calissendorff	  }\affiliation{Department of Astronomy, University of Michigan, 1085 S. University, Ann Arbor, MI 48103}	
\author[0000-0002-3968-3780]{Faustine Cantalloube    }\affiliation{Aix Marseille Univ, CNRS, CNES, LAM, Marseille, France}	
\author[0000-0001-7255-3251]{Gabriele Cugno	  }\affiliation{Department of Astronomy, University of Michigan, Ann Arbor, MI 48109, USA}
\author[0000-0003-1863-4960]{Matthew De Furio	  }\affiliation{Department of Astronomy, University of Michigan, Ann Arbor, MI 48109, USA}	
\author[0000-0001-9823-1445]{Trent J.~Dupuy	  }\affiliation{Institute for Astronomy, University of Edinburgh, Royal Observatory, Blackford Hill, Edinburgh, EH9 3HJ, UK}	
\author[0000-0002-8332-8516]{Samuel M.~Factor	  }\affiliation{Department of Astronomy, The University of Texas at Austin, Austin, TX 78712, USA}	
\author[0000-0001-6251-0573]{Jacqueline K.~Faherty  }\affiliation{Department of Astrophysics, American Museum of Natural History, Central Park West at 79th St., New York, NY 10024, USA}	
\author[0000-0002-0176-8973]{Michael P.~Fitzgerald  }\affiliation{University of California, Los Angeles, 430 Portola Plaza Box 951547, Los Angeles, CA 90095-1547}	
\author[0000-0003-4557-414X]{Kyle Franson	          }\affiliation{NSF Graduate Research Fellow}\affiliation{Department of Astronomy, The University of Texas at Austin, Austin, TX 78712, USA}	
\author[0000-0003-4636-6676]{Eileen C.~Gonzales	 }\affiliation{51 Pegasi b Fellow} \affiliation{Department of Astronomy and Carl Sagan Institute, Cornell University, 122 Sciences Drive, Ithaca, NY 14853, USA}	
\author[0000-0003-1150-7889]{Callie E.~Hood	}\affiliation{Department of Astronomy and Astrophysics, University of California, Santa Cruz, 1156 High St, Santa Cruz, CA 95064}
\author[0000-0002-4884-7150]{Alex R.~Howe	         }\affiliation{NASA Goddard Space Flight Center}	
\author[0000-0001-9811-568X]{Adam L.~Kraus	         }\affiliation{Department of Astronomy, The University of Texas at Austin, Austin, TX 78712, USA}	

\author[0000-0002-4677-9182]{Masayuki Kuzuhara	}\affiliation{Astrobiology Center of NINS, 2-21-1, Osawa, Mitaka, Tokyo, 181-8588, Japan}	\affiliation{National Astronomical Observatory of Japan, 2-21-2, Osawa, Mitaka, Tokyo 181-8588, Japan}	
\author{Anne-Marie Lagrange   }\affiliation{LESIA, Observatoire de Paris, Universit{\'e} PSL, CNRS, Universit{\'e} Paris Cit{\'e}, Sorbonne Universit{\'e}, 5 place Jules Janssen, 92195 Meudon, France}	
\author[0000-0002-6964-8732]{Kellen Lawson	        }\affiliation{NASA-Goddard Space Flight Center, 8800 Greenbelt Rd, Greenbelt, MD 20771, USA}	
\author[0000-0001-7819-9003]{Cecilia Lazzoni	}\affiliation{University of Exeter, School of Physics and Astronomy, Stocker Road, Exeter, EX4 4QL, UK}	
	
\author[0000-0001-7047-0874]{Pengyu Liu	        }\affiliation{SUPA, Institute for Astronomy, Royal Observatory, University of Edinburgh, Blackford Hill, Edinburgh EH9 3HJ, UK; Centre for Exoplanet Science, University of Edinburgh, Edinburgh EH9 3HJ, UK}	
\author[0000-0002-3414-784X]{Jorge Llop-Sayson	}\affiliation{Department of Astronomy, California Institute of Technology, Pasadena, CA 91125, USA}	
\author{James P.~Lloyd	}\affiliation{Department of Astronomy, Cornell University, Ithaca NY 14850, USA}	
\author[0000-0001-6301-896X]{Raquel A.~Martinez	}\affiliation{ Department of Physics and Astronomy, University of California, Irvine, 4129 Frederick Reines Hall, Irvine, CA 92697-4575}
\author[0000-0002-9133-3091]{Johan Mazoyer	         }\affiliation{LESIA, Observatoire de Paris, Universit{\'e} PSL, CNRS, Universit{\'e} Paris Cit{\'e}, Sorbonne Universit{\'e}, 5 place Jules Janssen, 92195 Meudon, France}	
\author[0000-0003-3829-7412]{Sascha P.~Quanz	}\affiliation{ETH Zurich, Institute for Particle Physics \& Astrophysics, Wolfgang-Pauli-Str. 27, 8093 Zurich, Switzerland}	
\author[0000-0002-4489-3168]{Jea Adams Redai	}\affiliation{Center for Astrophysics ${\rm \mid}$ Harvard {\rm \&} Smithsonian, 60 Garden Street, Cambridge, MA 02138, USA}	
\author[0000-0001-9992-4067]{Matthias Samland	}\affiliation{Max-Planck-Institut f\"ur Astronomie, K\"onigstuhl 17, 69117 Heidelberg, Germany}	
\author[0000-0001-5347-7062]{Joshua E.~Schlieder   }\affiliation{NASA Goddard Space Flight Center, Greenbelt, MD 20771, USA}	
\author[0000-0002-6510-0681]{Motohide Tamura	}\affiliation{The University of Tokyo, 7-3-1 Hongo, Bunkyo-ku, Tokyo 113-0033, Japan}	
\author[0000-0003-2278-6932]{Xianyu Tan	        }\affiliation{Atmospheric, Ocean, and Planetary Physics, Department of Physics, University of Oxford, UK}	
\author[0000-0002-6879-3030]{Taichi Uyama	}\affiliation{Infrared Processing and Analysis Center, California Institute of Technology, 1200 E. California Blvd., Pasadena, CA 91125, USA}	
\author[0000-0002-5902-7828]{Arthur Vigan	}\affiliation{Aix Marseille Univ, CNRS, CNES, LAM, Marseille, France}	
\author[0000-0003-0489-1528]{Johanna M.~Vos	}\affiliation{Department of Astrophysics, American Museum of Natural History, Central Park West at 79th Street, NY 10024, USA}	
\author[0000-0002-4309-6343]{Kevin Wagner	}\affiliation{NASA Hubble Fellowship Program – Sagan Fellow}\affiliation{Steward Observatory, University of Arizona, 933 N. Cherry Ave., Tucson, AZ 85721, USA}	
\author[0000-0002-9977-8255]{Schuyler G.~Wolff	}\affiliation{Steward Observatory, University of Arizona, 933 N. Cherry Ave., Tucson, AZ 85721, USA}	
\author[0000-0001-7591-2731]{Marie Ygouf	}\affiliation{Jet Propulsion Laboratory, California Institute of Technology, Mail Stop 321-100, 4800 Oak Grove Drive, Pasadena CA 91109  USA}	
\author{Xi Zhang	        }\affiliation{Department of Astronomy and Astrophysics, University of California, Santa Cruz, 1156 High St, Santa Cruz, CA 95064}	
\author[0000-0002-9870-5695]{Keming Zhang	}\affiliation{Department of Astronomy, 501 Campbell Hall, University of California Berkeley, Berkeley, CA 94720-3411, USA}
\author[0000-0002-3726-4881]{Zhoujian Zhang} \affiliation{Department of Astronomy, The University of Texas at Austin, Austin, TX 78712, USA}



\begin{abstract}
We present the highest fidelity spectrum to date of a planetary-mass object. VHS 1256 b is a $<$20 M$_\mathrm{Jup}$ widely separated ($\sim$8\arcsec, a = 150 au), young, planetary-mass companion that shares photometric colors and spectroscopic features with the directly imaged exoplanets HR 8799 c, d, and e. As an L-to-T transition object, VHS 1256 b exists along the region of the color-magnitude diagram where substellar atmospheres transition from cloudy to clear. We observed VHS 1256~b with \textit{JWST}'s NIRSpec IFU and MIRI MRS modes for coverage from 1 $\mu$m to 20 $\mu$m at resolutions of $\sim$1,000 - 3,700. Water, methane, carbon monoxide, carbon dioxide, sodium, and potassium are observed in several portions of the \textit{JWST} spectrum based on comparisons from template brown dwarf spectra, molecular opacities, and atmospheric models. The spectral shape of VHS 1256 b is influenced by disequilibrium chemistry and clouds. We directly detect silicate clouds, the first such detection reported for a planetary-mass companion.
\end{abstract}



\section{Introduction}
\label{sec:intro}

The light observed from an exoplanet contains information about the planet's composition, atmospheric dynamics, and other bulk physical properties.  This, in turn, can be used to infer how the planet formed and evolved.  Various parts of the planet's spectrum contain different information \citep[e.g.][]{1997ApJ...491..856B,2005ARA&A..43..195K}.  For example, in a $\sim$1000K gas-giant exoplanet or a more massive brown dwarf analog, the visible part of the spectrum ($<$1 $\mu$m) contains alkali lines that can constrain metallicity and surface gravity \citep{2003ApJ...594..510B}, the near-infrared part of the spectrum (1-5 $\mu$m) contains water (H$_{2}$O), carbon monoxide (CO), and methane (CH$_{4}$) absorption bands that can constrain atomic ratios (e.g., C/O, C/H, O/H) and turbulent mixing \citep{2011ApJ...733...65B,2013Sci...339.1398K}, and the mid-infrared part of the spectrum ($>$5 $\mu$m) contains a solid state silicate feature that can be used to measure the compositions of clouds \citep{2005ApJ...623.1115C,2022MNRAS.513.5701S}.  The full wavelength range constrains the temperature-pressure profile of the atmosphere and the muting and reddening effects of clouds \citep{2013cctp.book..367M}.  

\textit{JWST} provides our first opportunity to explore the spectra of brown dwarfs and exoplanets over their full luminous range \citep{2022arXiv220705632R}.  Previously, the longest wavelength exoplanet images available were taken at 5 $\mu$m \citep[e.g.][]{2010ApJ...716..417H,2011ApJ...739L..41G,2015ApJ...815..108M,2017AJ....154...10R}, while eclipse measurements of transiting planets have occasionally extended to photomtry and spectroscopy out to 24 $\mu$m \citep[e.g.][]{2008ApJ...673..526K,2008ApJ...686.1341C,2008Natur.456..767G}.  \textit{JWST} can measure exoplanet spectral energy distributions (SEDs) with multi-wavelength photometry and spectroscopy from 0.6-28.1 $\mu$m, a range that contains essentially 100\% of the energy emitted by a 300 K exoplanet and 99.6\% of the energy emitted by a 1000 K exoplanet \citep[based~on~models~from][]{Morley14, 2021ApJ...920...85M}.

The Early Release Science (ERS) Program, \textit{High Contrast Imaging of Exoplanets and Exoplanetary Systems with JWST} (ERS 1386, PI Hinkley),  
employs several different modes of \textit{JWST} appropriate for studying
directly imaged exoplanets, planetary-mass companions, and the circumstellar disks in which they form \citep{2022arXiv220512972H}.  From a technical perspective, the program has been designed to assess the performance of the Near-Infrared Camera (NIRCam) and Mid-Infrared Instrument (MIRI) coronagraphic modes as well as the Near InfraRed Imager and Slitless Spectrograph (NIRISS) aperture masking interferometry mode with the goal of preparing the community for ambitious direct imaging surveys in Cycle 2 and beyond.  

Beyond understanding \textit{JWST}'s technical performance, the program also aims to provide template spectra of exoplanet atmospheres over \textit{JWST}'s full wavelength range and improve our understanding of gas giant atmospheric physics and chemistry \citep{2022arXiv220512972H}.  In the high-contrast regime, \textit{JWST} uses coronagraphs in NIRCam and MIRI to suppress light from an exoplanet's host star \citep{2005SPIE.5905..185G,2009SPIE.7440E..0WK,2015PASP..127..633B, 2022arXiv220711080B}.  Both of these modes are imaging only---\textit{JWST}'s spectroscopic modes do not work with coronagraphy.  However, in order to provide the best possible spectral template observations, the \textit{High Contrast Imaging of Exoplanets and Exoplanetary Systems with JWST} program elected to use the Near-Infrared Spectrograph (NIRSpec) and MIRI integral field spectrographs \citep{2022A&A...661A..82B, 2015PASP..127..646W} to observe a widely separated planetary-mass companion \footnote{As discussed in \citet{2022arXiv220808448D} and later in this paper, VHS 1256 b’s mass is either just above or just below the deuterium burning limit that is often used as a dividing line between exoplanets and brown dwarfs.  Its formation mechanism is unknown, although its wide orbital separation suggests that it did not form like a typical exoplanet.  Depending on its true mass and other context, VHS 1256 b could correctly be referred to as an exoplanet or brown dwarf.  Similar to \citet{2022arXiv220808448D}, we adopt the intermediate term, \textit{planetary-mass companion}.}, where coronagraphy is not necessary: VHS J125601.92-125723.9 b (hereafter VHS 1256 b).  When selecting targets, our team searched for an object with the following characteristics: (1) a planet-like spectrum, (2) a wide separation and low contrast with its host star so that it could be observed with integral field spectroscopy instead of coronagraphic imaging, and (3) no overlap with the Guaranteed Time Observation (GTO) programs.Among the potential targets considered, VHS 1256 b best fulfilled these requirements.

\begin{figure}
    \centering
    \includegraphics[trim={0 0 .16cm 0}, width = 3 in]{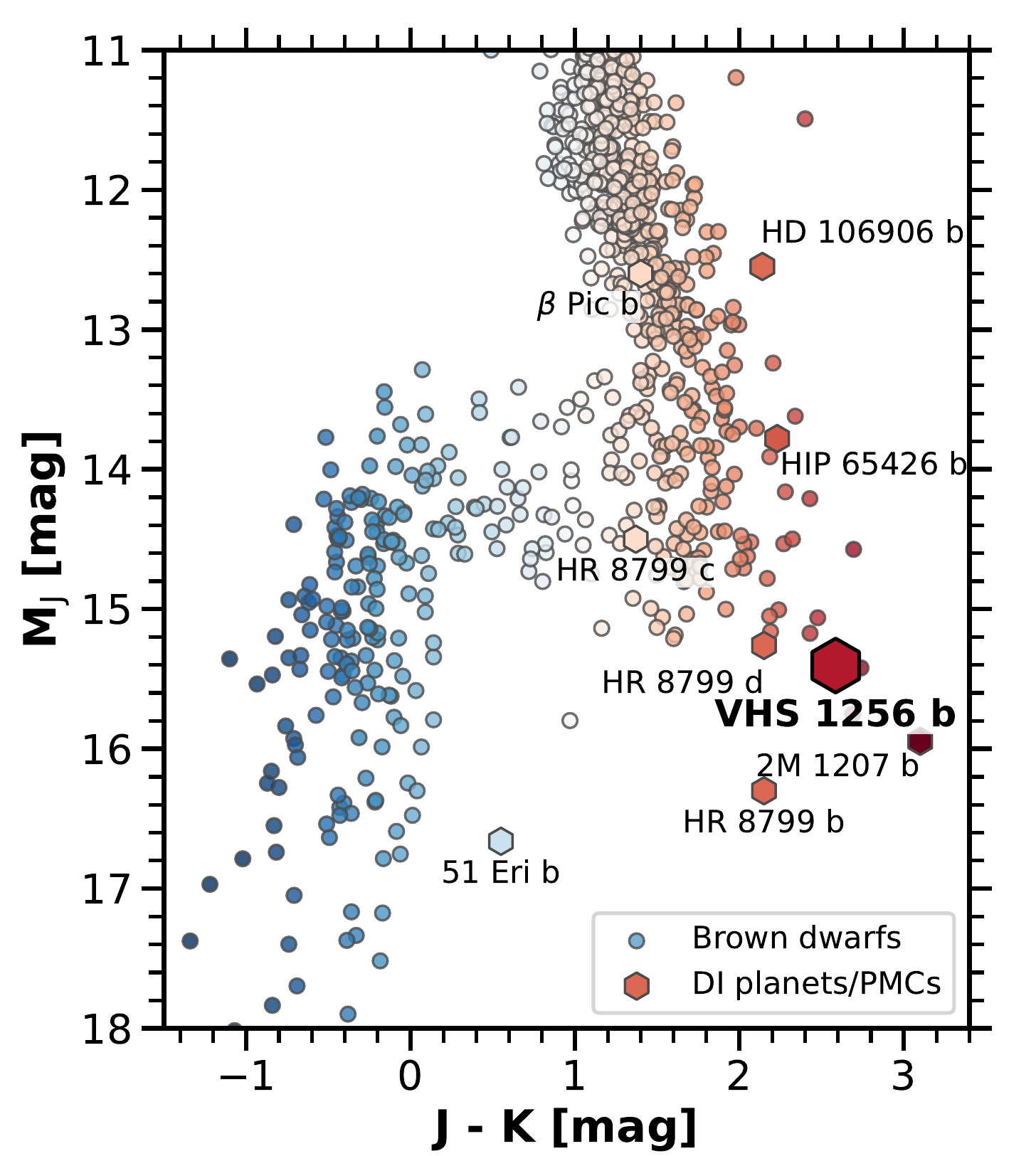}
    \caption{VHS 1256 b (red hexagon) is an analog to directly imaged exoplanets like the HR 8799 planets due to similar near-infrared colors. VHS 1256 b and other directly imaged exoplanets maintain these colors compared to similar temperature brown dwarfs because of lower surface gravities, which allow clouds to linger in the upper atmosphere. Photometry from \url{http://bit.ly/UltracoolSheet}}.
    \label{fig:CMD}
\end{figure}

VHS 1256 b was discovered by \citet{2015ApJ...804...96G} $\sim$8" ($\sim$150 au) away from an M-dwarf binary \citep{2016ApJ...818L..12S,2016ApJ...830..114R}. Spectra of the host M-star binary and the planetary-mass companion are both consistent with youth ($\lesssim$300 Myr); the companion has red colors and a triangular H-band feature that is seen in other low-mass companions, as well as weak alkali lines \citep{2015ApJ...804...96G,2022arXiv220706622P}. The photometry \citep{2015ApJ...804...96G} and distance \citep[22.2 pc;][]{2020RNAAS...4...54D} of VHS 1256 b place it in the vicinity of other faint red directly imaged planets, such as HR 8799 cde \citep{2008Sci...322.1348M,2010Natur.468.1080M} and 2MASS 1207 b \citep{2004A&A...425L..29C} in a color-magnitude diagram (see Figure~\ref{fig:CMD}). This distribution of redder photometric colors are theorized to be the result of cloudy atmospheres \citep{2011ApJ...737...34M,2011ApJ...729..128C,2011ApJ...732..107S,2012ApJ...753...14S,2012ApJ...754..135M}, which can linger in the upper atmospheres of young brown dwarfs, due to low surface gravities being prevalent among the lowest-mass young brown dwarfs \citep{2016ApJ...833...96L,2016ApJS..225...10F}.

Spectroscopic, near-infrared studies of VHS 1256 b show absorption from H$_{2}$O and CO like other L-to-T transition brown dwarfs, but no CH$_{4}$ despite cool effective temperatures that could produce that molecule \citep{2022arXiv220706622P, 2022arXiv220703819H}. Additionally, VHS 1256 b's L-band spectrum shows weaker absorption compared to older field brown dwarfs \citep{2018ApJ...869...18M}, which is an indication of disequilibrium chemistry that is also apparent in the spectral energy distributions of the HR 8799 planets \citep{2011ApJ...733...65B,2014ApJ...792...17S}.  VHS 1256 b does not belong to a known moving group \citep{2015ApJ...804...96G,2020RNAAS...4...54D}, and therefore, at the time it was selected for the \textit{JWST} program, its age and mass were uncertain.  More recently, \citet{2022arXiv220808448D} have determined a system age of 140$\pm$20 Myr, based on the measured dynamical mass and bolometric luminosity of the inner binary combined with \citet{2015A&A...577A..42B} evolutionary models, confirming VHS 1256 b's youth and $<$20 $M_{\mathrm{Jup}}$ mass.

After VHS 1256 b was chosen as an ERS target, \citet{2020ApJ...893L..30B} discovered that it has the largest known variability amplitude of any L dwarf. Later, \cite{2022arXiv221002464Z} determined that VHS 1256 b has the largest known variability amplitude of any brown dwarf (38\% peak-to-peak at J-band, separated by 2 years).  This variability is potentially caused by heterogeneous temperature and cloud distributions that may be particularly prevalent in the atmospheres of young, low-mass planets \citep{2013ApJ...768..121A, 2017Sci...357..683A, 2019MNRAS.483..480V, 2020AJ....160...77Z, 2021ApJ...906...64A, 2022BAAS...54e.206V}.  Therefore, our team has embarked on a campaign to study the object's variability with ground-based telescopes, which will be published in a subsequent paper. Some of our variability observations are nearly contemporaneous with the \textit{JWST} ERS 1386 NIRSpec and MIRI observations.  Based on these observations, in Section~\ref{sec:discussion} we constrain the impact of variability on the flux calibration of our spectra during the epoch of our \textit{JWST} observations.

In this paper, we analyze the ERS 1386 observations of VHS 1256 b, spanning 0.97 $\mu$m - 28.1 $\mu$m.  Due to \textit{JWST}'s broad wavelength coverage, spectacular sensitivity, and freedom from telluric absorption, these data are the highest quality ever published for an exoplanet or brown dwarf to date.  In Section~\ref{sec:observations}, we describe the \textit{JWST} observations. In Section~\ref{sec:reduction}, we describe our data reduction and present a 0.97 $\mu$m-19.8 $\mu$m spectrum of VHS 1256 b.  In Section~\ref{sec:molecular features} we indentify features from molecular absorption due to gaseous species and from cloud species. In Section~\ref{sec:modeling}, we perform forward model comparisons to understand atmospheric properties such as disequilibrium chemistry driven by atmospheric mixing and the influence of clouds on VHS 1256 b's spectral energy distribution. Lastly, in Section~\ref{sec:discussion} we discuss the object's luminosity, the effect of VHS 1256 b's intrinsic variability on our data, and the impact of the \textit{JWST} spectrum on our understanding of the properties of directly imaged exoplanets.

\section{Observations}
\label{sec:observations}
Spectroscopic observations of VHS 1256 b were obtained using NIRSpec and MIRI on \textit{JWST}. NIRSpec observations were acquired from UT 09:43:42 to 11:48:08 on 2022-07-05.  NIRSpec was used in IFU mode to measure a spectrum of VHS 1256 b between 0.97 and 5.27 $\mu$m at resolutions of  $\sim$1000 - 2700,  with the following filter/grating combinations: G140H/F100LP, G235H/F170LP, G395H/F290LP. Each NIRSpec observation with a given filter/grating combination used the full detector with the NRSIRS2RAPID readout and a 4-point dither box pattern (0.4\arcsec on a side), for a total exposure time of 2144.57 seconds per filter/grating combination. Over about 1.5 hours, the sequential order of the NIRSpec observation modes were G235H/F170LP, G395H/F290LP, followed by G140H/F100LP with 9 minutes of downtime between each observation mode. MIRI observations were acquired from UT 11:56:54 to 13:56:05 on 2022-07-05. The MIRI observations of VHS 1256 b were completed in the short, medium, and long grating settings of all 4 IFU channels for overlapping coverage from 4.98 to 28.1 microns with a resolution of $\sim$1300 - 3000 using 4-point dithering. We obtained an observation in each MIRI grating/channel mode with the FASTR1 read out setting and total exposure times of 1576.22 seconds. Channel 1 and 2 had their short, medium, and long observations taken first, then Channel 3 and 4 using the short, medium, and long gratings. An overview of the observations taken are provided in Table~\ref{tab:observations} and~\ref{tab:bg-observations}. 

\begin{deluxetable*}{llcllcc}
\tablenum{1}
\tablecaption{Observations of VHS 1256 b with \textit{JWST} \label{tab:observations}}
\tablewidth{0pt}
\tablehead{
\colhead{Instrument} & \colhead{Mode} & \colhead{Wavelength} & \colhead{Subarray} & \colhead{Readout} &  \colhead{Resolving Power} & \colhead{Exposure Time (s)} }
\startdata
NIRSpec & G140H/F100LP         & 0.97 - 1.89   & FULL & NRSIRS2RAPID & $\sim$1000 & 1283.82  \\
NIRSpec & G235H/F170LP         & 1.66 - 3.17   & FULL & NRSIRS2RAPID & $\sim$2700 &1283.82 \\
NIRSpec	& G395H/F290LP         & 2.87 - 5.27   & FULL & NRSIRS2RAPID & $\sim$2700 &1283.82 \\
MIRI    & Channel 1, Short A   & 4.9 – 5.74    & FULL & FASTR1       & 3,320 – 3,710 & 1576.22\\
MIRI    & Channel 1, Medium B  & 5.65 - 6.63   & FULL & FASTR1         & 3,190 – 3,750 & 1576.22\\
MIRI    & Channel 1, Long C    & 6.53 - 7.65   & FULL & FASTR1         & 3,100 – 3,610 & 1576.22\\
MIRI    & Channel 2, Short A   & 7.51 - 8.76   & FULL & FASTR1         & 2,990 – 3,110 & 1576.22 \\
MIRI    & Channel 2, Medium B  & 8.67 - 10.15  & FULL & FASTR1         & 2,750 – 3,170 & 1576.22\\
MIRI    & Channel 2, Long C    & 10.01 - 11.71  & FULL & FASTR1         & 2,860 – 3,300 & 1576.22\\
MIRI    & Channel 3, Short A   & 11.55 - 13.47 & FULL & FASTR1         & 2,530 – 2,880 & 1576.22\\
MIRI    & Channel 3, Medium B  & 13.29 - 15.52 & FULL & FASTR1         & 1,790 - 2,640 & 1576.22\\
MIRI    & Channel 3, Long C    & 15.41 - 18.02 & FULL & FASTR1         & 1,980 – 2,790 & 1576.22\\
MIRI    & Channel 4, Short A   & 17.71 - 20.94 & FULL & FASTR1         & 1,460 – 1,930 & 1576.22 \\
MIRI    & Channel 4, Medium B  & 20.69 - 24.44  & FULL & FASTR1         & 1,680 – 1,770 & 1576.22\\
MIRI    & Channel 4, Long C    & 23.22 - 28.1  & FULL & FASTR1        & 1,630 – 1,330 & 1576.22\\
\enddata
\tablecomments{Breakdown of the early release science observations of VHS 1256 b completed for our program. Both JWST/NIRSpec and JWST/MIRI utilized 4 point dithering over the exposure. Resolution values for each of the instrument modes are from the following websites: \\
NIRSpec:\url{https://jwst-docs.stsci.edu/jwst-near-infrared-spectrograph/NIRSpec-observing-modes/NIRSpec-ifu-spectroscopy} \\
MIRI: \url{https://jwst-docs.stsci.edu/jwst-mid-infrared-instrument/miri-observing-modes/miri-medium-resolution-spectroscopy}}
\end{deluxetable*}

\begin{deluxetable*}{llcllcc}
\tablenum{2}
\tablecaption{MIRI Background Observations \label{tab:bg-observations}}
\tablewidth{0pt}
\tablehead{
\colhead{Instrument} & \colhead{Mode} & \colhead{Wavelength} & \colhead{Subarray} & \colhead{Readout} &  \colhead{Resolving Power} & \colhead{Exposure Time (s)} }
\startdata
MIRI    & Channel 1, Short A   & 4.9 – 5.74    & FULL & FASTR1       & 3,320 – 3,710 & 394.06\\
MIRI    & Channel 1, Medium B  & 5.65 - 6.63   & FULL & FASTR1       & 3,190 – 3,750 & 394.06\\
MIRI    & Channel 1, Long C    & 6.53 - 7.65   & FULL & FASTR1       & 3,100 – 3,610 & 394.06\\
MIRI    & Channel 2, Short A   & 7.51 - 8.76   & FULL & FASTR1       & 2,990 – 3,110 & 394.06\\
MIRI    & Channel 2, Medium B  & 8.67 - 10.15  & FULL & FASTR1       & 2,750 – 3,170 & 394.06\\
MIRI    & Channel 2, Long C    & 10.01 - 11.71  & FULL & FASTR1      & 2,860 – 3,300 & 394.06\\
MIRI    & Channel 3, Short A   & 11.55 - 13.47 & FULL & FASTR1       & 2,530 – 2,880 & 394.06\\
MIRI    & Channel 3, Medium B  & 13.29 - 15.52 & FULL & FASTR1       & 1,790 - 2,640 & 394.06\\
MIRI    & Channel 3, Long C    & 15.41 - 18.02 & FULL & FASTR1       & 1,980 – 2,790 & 394.06\\
MIRI    & Channel 4, Short A   & 17.71 - 20.94 & FULL & FASTR1       & 1,460 – 1,930 & 394.06\\
MIRI    & Channel 4, Medium B  & 20.69 - 24.44  & FULL & FASTR1      & 1,680 – 1,770 & 394.06\\
MIRI    & Channel 4, Long C    & 23.22 - 28.1  & FULL & FASTR1       & 1,630 – 1,330 & 394.06\\
\enddata
\tablecomments{Background observations taken with MIRI. The background observations were not used to reduce the VHS 1256 b spectra but are included in the summary of program observations. Resolution values for each of the instrument modes are from the following websites: \\
MIRI:\url{https://jwst-docs.stsci.edu/jwst-mid-infrared-instrument/miri-observing-modes/miri-medium-resolution-spectroscopy}}
\end{deluxetable*}

\section{Spectroscopic Reduction}
\label{sec:reduction}
\subsection{\textit{JWST}/NIRSpec}
Version 1.7.2 of the standard \textit{JWST} pipeline was used to reduce each science observation dither into spectral cubes for the NIRSpec IFU data set. The CRDS versions and context used by the JWST pipeline were `11.16.12' and `\texttt{jwst\_0977.pmap}' respectively. At the time of the initial analysis, the NIRSpec pipeline used instrument parameters determined from ground-based testing \citep{2022arXiv220705632R}. Residual bad pixels and cosmic rays remain in the standard pipeline reductions and these issues are compounded when dithers are reduced together. Each dither was processed through Stage 1 which performs detector-level corrections and converts detector ramps into slope images. After Stage 1, the Stage 2 step removes other instrument artifacts and creates calibrated slope images. Stage 3 of the pipeline takes slope images to create 3-dimensional spectral cubes of a target. Total errors are propagated through all stages of the JWST Pipeline starting with variances estimated in the slope fitting step in Stage 1. Stage 3 spectral cubes have an associated error array with the same dimensions of the spectral cube.

The cube building step in Stage 3 was run with outlier detection turned off and all parameters set to default values to produce 3-dimensional spectral cubes that are aligned in right ascension and declination. The Stage 3 spectral cubes were used to create 1-dimensional spectra using aperture extraction. First, the spectral cube for each dither is collapsed along the wavelength axis to calculate a mean image to mitigate the effects of cosmic rays, then a 2-D Gaussian is fit to the source point spread function (PSF) to find the source's center position. The center of the source's position changes on a subpixel level in both x and y directions of the image cube for all NIRSpec observation modes. The source center oscillates as a function of wavelength by .2-.3 pixels in the x-direction and .15-.2 pixels in the y-direction. For each image along the spectral cube wavelength axis, we then extract the flux within a circular aperture centered on this position. We tested several extraction radii to find where the object's flux plateaus with extraction radius. Based on this, we adopted extraction radii of 3-4 pixels (1.2 - 1.7 FWHM) for VHS 1256 b, while the calibrator star required a radius of 4 pixels across all bands. Aperture correction was not applied to the NIRSpec IFU data and it is not relevant for a well centered point source. The same aperture used for each image is applied to the error array for every wavelength to calculate the error with standard error propagation. The weighted mean of the dithers is taken to produce a final spectrum.

Before a final weighted average is calculated, a hard cutoff is applied to fluxes that are 10\% higher than the spectral energy distribution of the initial median to remove wavelengths affected by cosmic rays. Edges of the reconstructed IFU data cube include spatial elements which are not fully illuminated and the extracted spectra derived from these parts of the data cube are removed from the final spectrum. After the cutoffs are applied, we take the weighted mean of the dithers to obtain the final spectrum for a filter/grating combination. The error of the spectrum is the propagated error of the weighted average. 

As discussed previously, the flux calibration of the spectral cubes is not optimized due to the use of ground-based calibration files in the standard JWST pipeline. To produce a final flux-calibrated spectrum for VHS 1256 b, the extracted spectra must be multiplied by a scale factor, which is the ratio of a calibrated standard spectrum and the response it produces after processing with the JWST pipeline. The A3V star TYC 4433-1800-1, observed during commissioning (Program ID: 1128) has a calibrated flux reference in the CALSPEC Library \citep{2020AJ....160...21B} was used as a calibrator to adjust the spectral response of the VHS 1256 b extracted spectrum. 

All observation modes used in the ERS program were available for the calibrator star and we reduced these observations with the same \textit{JWST} pipeline parameters as the science spectra. As before, each dither was reduced individually and the extracted spectra of each filter/grating mode were combined with a weighted mean to produce a final spectrum for that mode. There are oscillating features present in both the calibrator star and the VHS 1256b NIRSpec spectra that are dependent on dither position and brightness. These features are averaged out in the calibrator star by fitting a fourth-order ($\sim$T$^{4}$) polynomial to obtain the ``mean" response at each wavelength.The intrinsic spectral shape of the planetary mass companion is not assumed and the oscillations are not removed by fitting. The oscillating features change the amplitude of the extracted science spectra by about 2 - 6\% however, in the extracted calibrator spectra these amplitudes can vary by as much as 20\% making the fitting step much more important for the calibrator star. In the VHS 1256b extracted spectra, the oscillations are visually apparent in dithers from the G140H/F100LP detector 1, G140H/F100LP detector 2, G140H/F100LP detector1, and G140H/F100LP detector 2 observations. Averaging or taking the mean of several dithers helps to remove some of these oscillating features, but there are not enough dithers to remove the effect entirely. The calibrator star has oscillations in all observation modes for our program. Portions of the final VHS 1256b spectra that appear to be affected by these oscillations are 1.65~$\mu$m - 1.75~$\mu$m and 2 $\mu$m - 2.4 $\mu$m. We will discuss this in context to identified molecular features in Section~\ref{sec:molecular features}. Generally speaking the crest to crest width of the oscillations appearing in the calibrator star extracted spectra vary in width, but are never smaller than .01 $\mu$m and can be as large as .029$\mu$m.

The standard spectrum of our calibrator star possesses absorption features which are masked before fitting a fourth-order polynomial to the spectra. We perform each fit for a single filter/grating mode. The best fit of the calibrator spectrum and the best fit of the extracted pipeline spectra are divided to find the scale factor, which is then applied to the VHS 1256 b spectrum that falls within that bandpass. The extracted errors from the calibrator spectra are also included with VHS 1256b's spectrum errors when the scale factor is applied.

\subsection{\textit{JWST}/MIRI}
Version 1.8.1 of the standard \textit{JWST} pipeline was used to reduce the MIRI MRS observation listed in Table~\ref{tab:observations}. The CRDS versions and context used by the JWST pipeline were `11.16.14' and `\texttt{jwst\_1007.pmap}' Each dithering sequence within a wavelength bandpass was combined to create a single spectral cube. \textit{JWST} pipeline stages 1, 2, and 3 were all used to reduce the MIRI MRS data.  The MIRI MRS dithers were reduced together for each channel/grating combination and 1-dimensional spectra were extracted from the resulting spectral cubes using aperture photometry.

We chose not to use the background subtraction method implemented in the standard pipeline, as it was unable to account for cosmic ray showers present in the data.  These cosmic ray showers appear as diffuse, extended structures that do not produce the usual jumps in the detector ramps, which the pipeline is able to detect. The distribution and shape of the cosmic ray showers vary between science and background exposures, and the impact of these differences produce the main systematic noise source for the faint point source extracted with MIRI.  We instead estimated the background using  a reference aperture placed off of the target in our calibrated science cubes. The separation of the background apertures varied between channels, with 1.7 arcsec for Channel 1, 2.2 arcsec for Channel 2, and 2.3 arcsec for Channels 3 and 4. 

Before performing aperture photometry, as with NIRSpec, each spectral cube is collapsed in the wavelength direction using the mean, and the source center is fit with a 2-D Gaussian. This is a necessary step due to slight band-to-band offsets still present in the distortion model of the MRS.  We then extracted a flux within a circular aperture centered on our derived source position for each image along the spectral cube wavelength axis.  For a given wavelength dependent extraction radius (2.0 FWHM) the aperture correction is applied to account for flux of the PSF missed outside of extraction radius. This correction factor was derived using webbPSF \citep{2014SPIE.9143E..3XP} models for the MRS, matching the empirical PSF FWHM measured during commissioning.

After extraction and aperture correction, fringes remain in portions of the MIRI spectrum. Fringing appears as a regular, beating interference pattern in the MIRI spectra \citep{2020A&A...641A.150A}, particularly in channels 2C, 3A, 3B, and 3C. The \texttt{ResidualFringeStep} is applied to the 1-dimensional spectra to find and remove fringing patterns from the data (Kavanagh et al. in prep.)\footnote{This step is not automatically applied on the 1d spectra by the \textit{JWST} pipeline at the moment. For more information see \url{https://jwst-docs.stsci.edu/jwst-calibration-pipeline-caveats/jwst-miri-mrs-pipeline-caveats}}.

Past 18 $\mu$m, the sensitivity of the MRS drops significantly due to a combination of low efficiency of the grating, low quantum efficiency of the detectors, and rising thermal background \citep{2015PASP..127..686G}. As a result the spectrum could not reliably be extracted in channel 4.  Instead, the cube was collapsed from 17.71 to 20.94 $\mu$m (Channel 4A) to produce a single photometric point. The source center is found by fitting a 2-D Gaussian to the collapsed channel 4A cube. A circular aperture (radius of 1.5 FWHM) is placed at the measured PSF center and the same aperture correction described above is used. We measure the background using an off-target reference aperture with the same radius and subtract the background.

The error in the extracted spectrum was estimated using the propagated error by the \texttt{jwst} pipeline \texttt{Extract1dStep}. The output x1d.fits files contain an ERR array extension that captures the error of the full processing chain, including the photon noise of the source. The extracted spectrum of the pipeline uses an annulus subtraction that introduces systematic errors form the background subtraction, but the overall spectrum matches the flux and shape of our own extraction. We therefore assume that the error reported by the pipeline should be representative for our own extraction. Indeed, the slope of the error follows the expected rising thermal background, but seems to be overall a factor of ten lower than what one would expect from the ETC. Additionally, comparing with the noise from a collection of reference apertures around the field of view confirms this discrepancy. We therefore choose to artificially amplify the pipeline estimated error by a factor of 10, yielding a rather conservative estimate of the S/N until the pipeline issue is resolved.

\begin{figure*}
    \centering
    \includegraphics[width = 6.68 in]{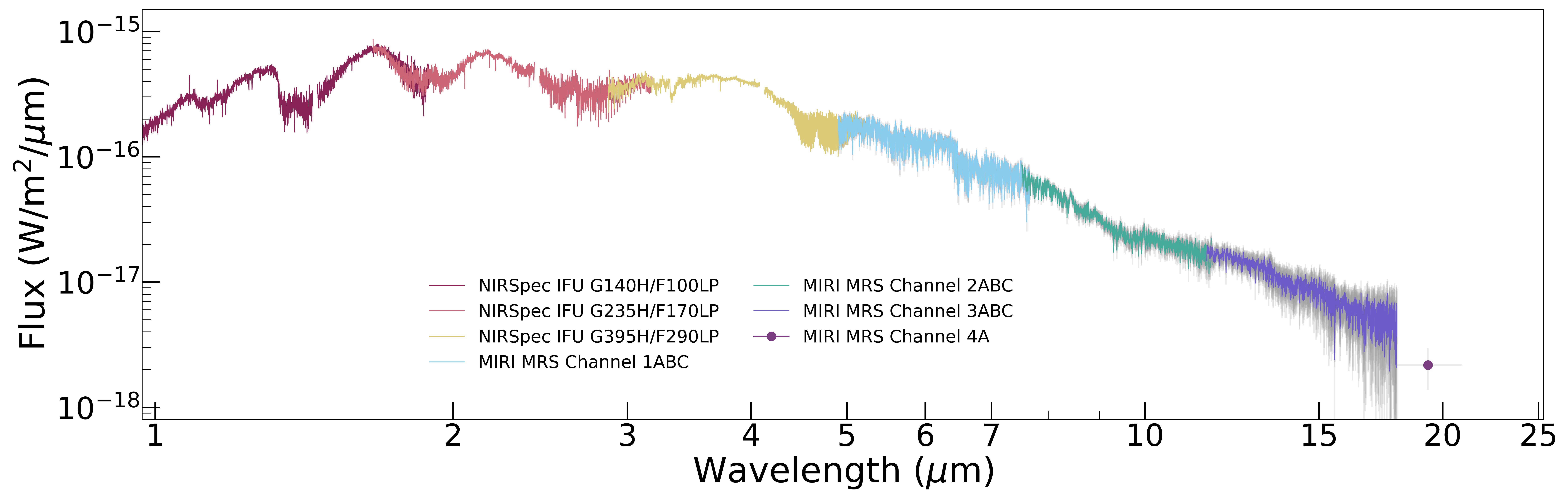}
    \caption{The full spectrum of VHS 1256 b using \textit{JWST}'s NIRSpec IFU and MIRI MRS observation modes. Bandpasses are highlighted with different colors and error bar are displayed in grey. A single photometric point for MIRI MRS Channel 4A is shown because there is little to no signal in the MIRI MRS 4B, and 4C channels. Error bars are plotted in a light grey.}
    \label{fig:full spectrum}
\end{figure*}

The full reduced spectrum with labeled bandpasses is shown in Figure~\ref{fig:full spectrum}. The NIRSpec (1 $\mu$m - 5 $\mu$m)  spectra have a signal to noise of 50 - 400 per pixel. The MIRI (4.9 $\mu$m - 20.94 $\mu$m) spectra have a signal-to-noise of 7 to 20 per pixel, with hardly any signal in Channel 4 (17.66 $\mu$m - 28.1 $\mu$m). The photometry point from channel has a signal-to-noise of 2.7.

\subsection{NIRSpec-MIRI Instrument Overlap} 
\begin{figure*}
    \centering
    \includegraphics[width = 6.5 in]{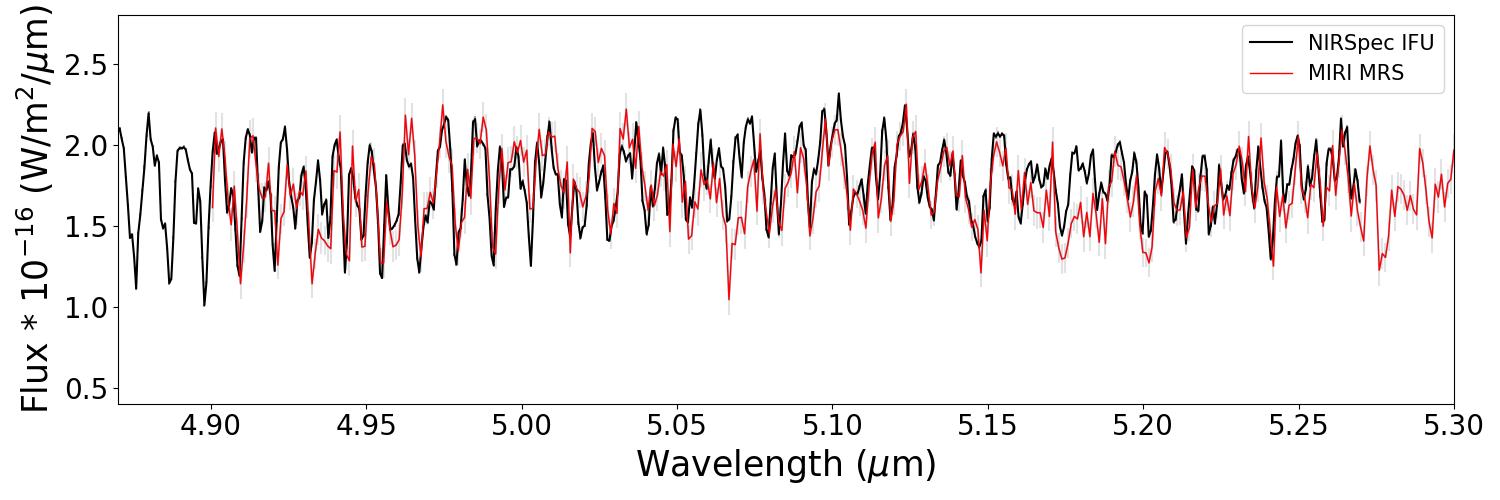}
    \caption{The wavelength overlap region between \textit{JWST}'s NIRSpec IFU (black) and MIRI MRS (red) observation modes with errors plotted in light gray. The flux calibrated NIRSpec IFU spectra agree well with the MIRI MRS extracted spectra in the overlap region. }
    \label{fig:overlap}
\end{figure*}

There are 557 data points from NIRSpec and 370 data points from MIRI that overlap in wavelength. When the NIRSpec data are binned down and interpolated onto MIRI’s wavelength spacing some portions of the NIRSpec data have a higher baseline flux, more than 3-sigma away from the MIRI data points. In Figure~\ref{fig:overlap} shows the overlap region of NIRSpec IFU and MIRI MRS data at the resolution of each respective instrument.

\section{Atmospheric Features}
\label{sec:molecular features}
\begin{figure*}
    \centering
    \includegraphics[width = 7 in]{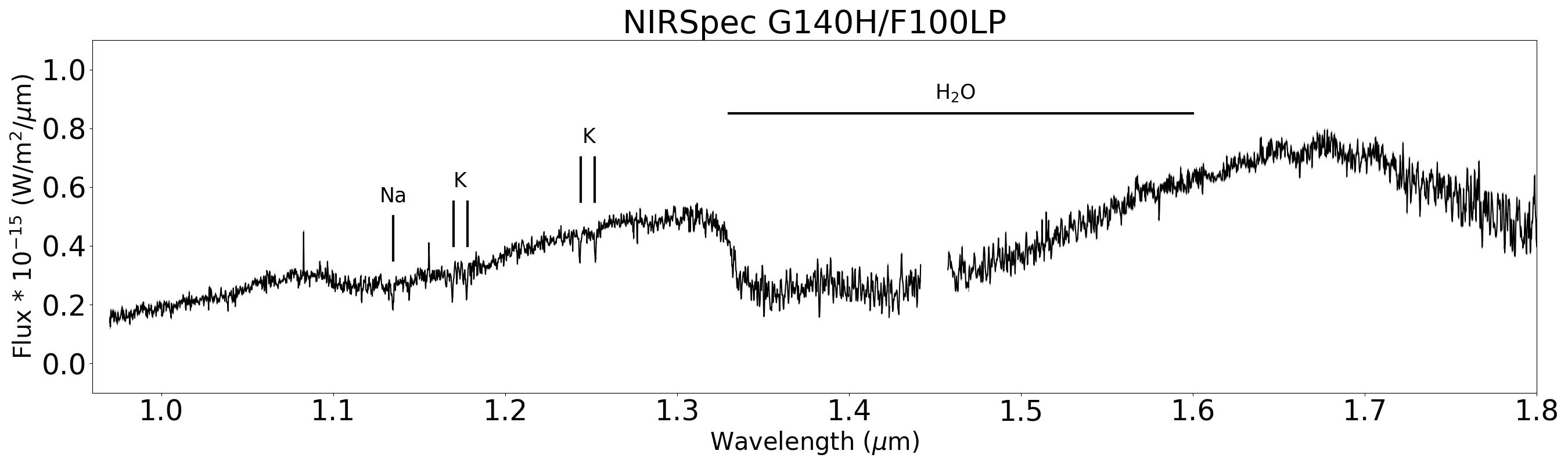}
    \includegraphics[width = 7 in]{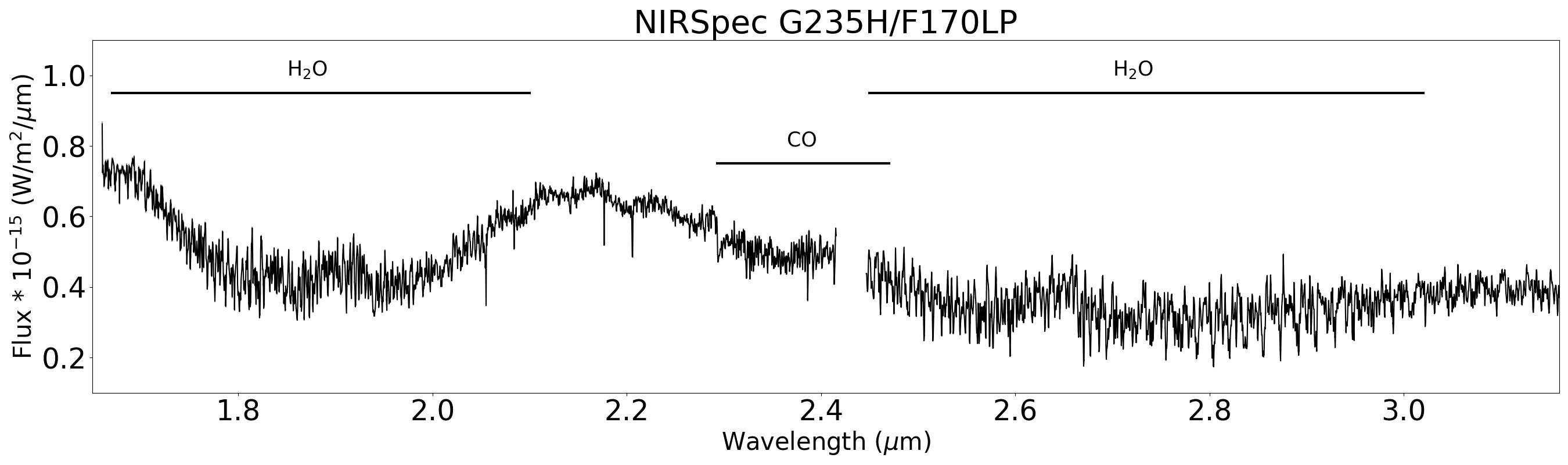}
    \includegraphics[width = 7 in]{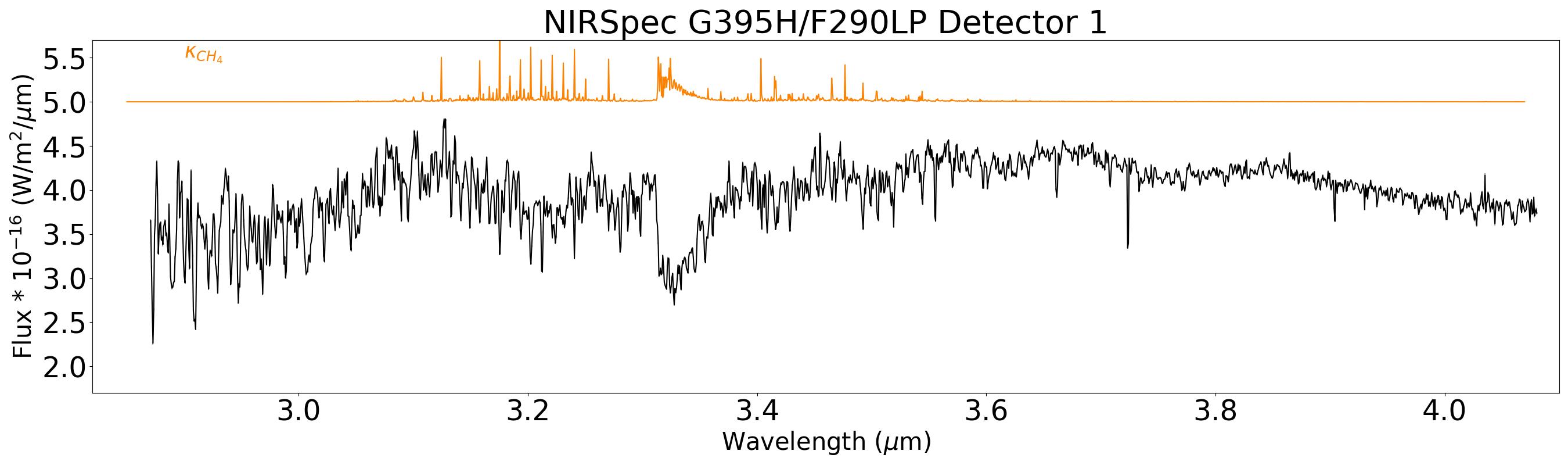}
    \includegraphics[width = 7 in]{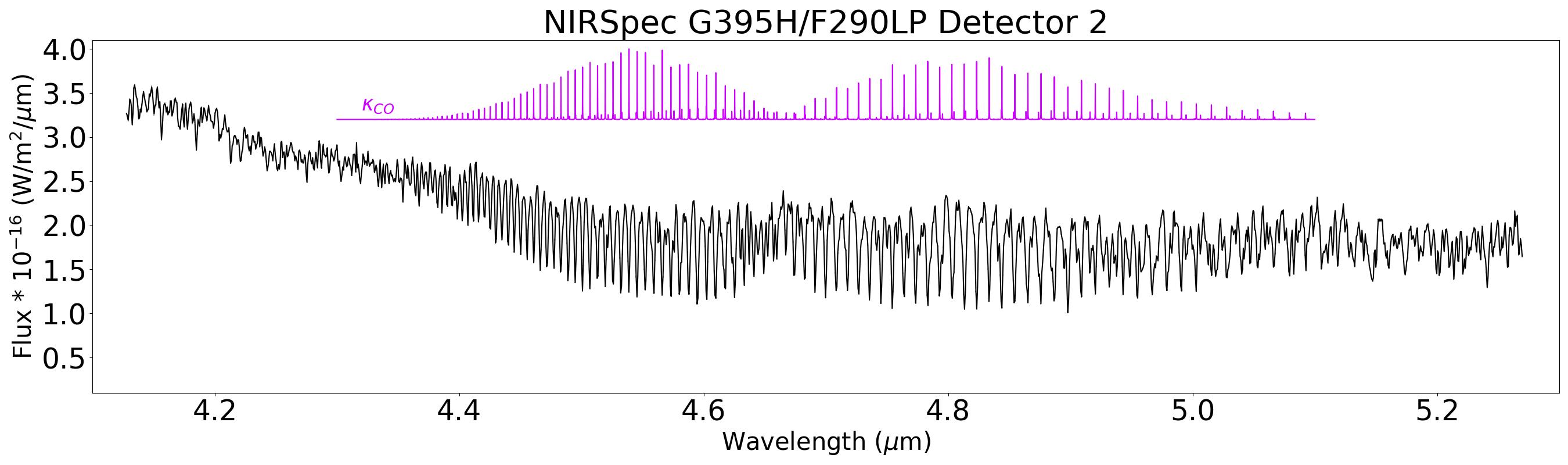}
    \caption{\textit{JWST}/NIRSpec spectrum of  VHS 1256 b, with important molecular gases highlighted. 
    Molecules were identified via visual comparison with template spectra.}
    \label{fig:NIRSpec Features}
\end{figure*}

\begin{figure*}
    \centering
    \includegraphics[width = 6 in]{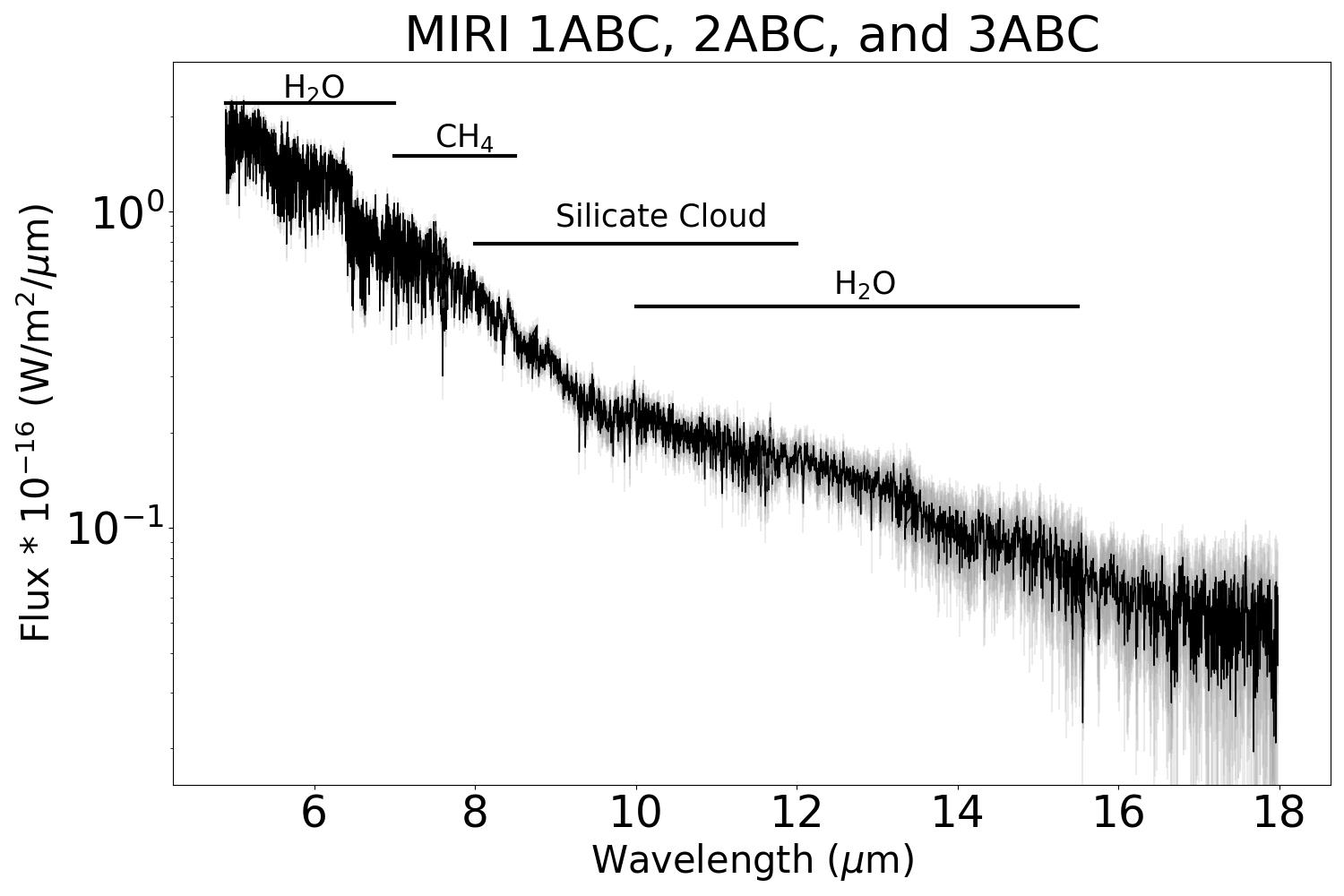}
    \includegraphics[width = 6 in]{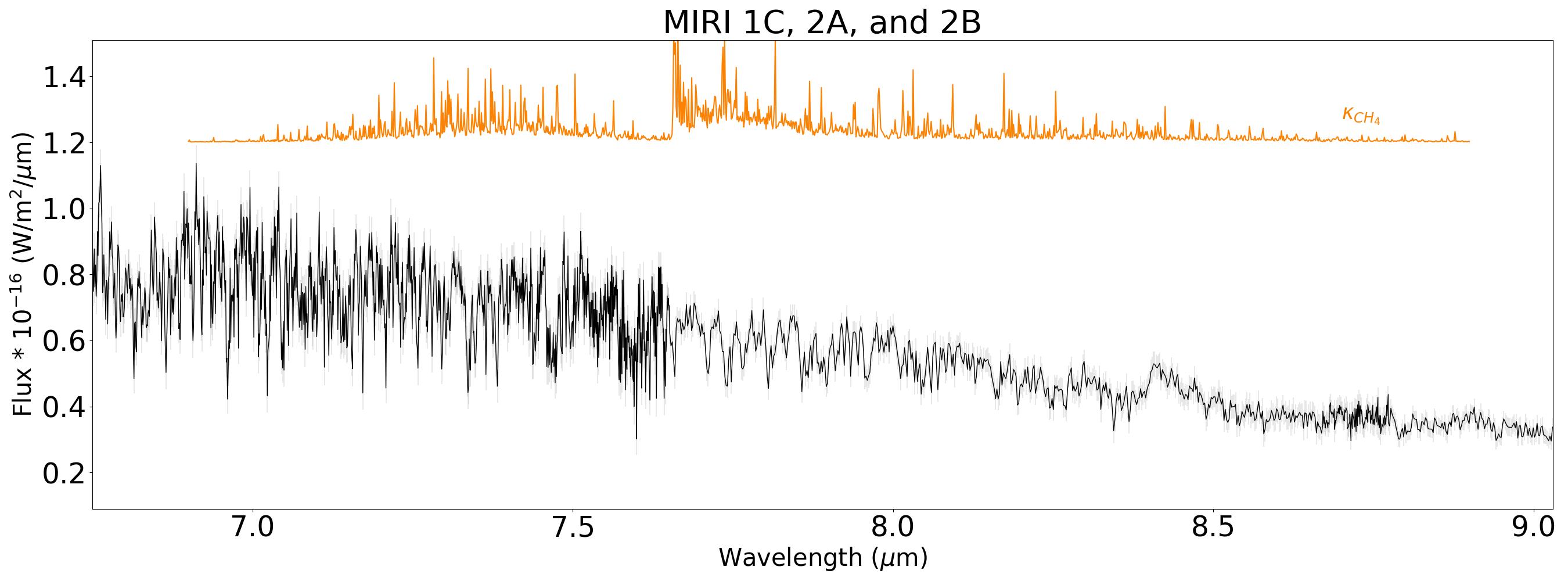}
    \includegraphics[width = 6 in]{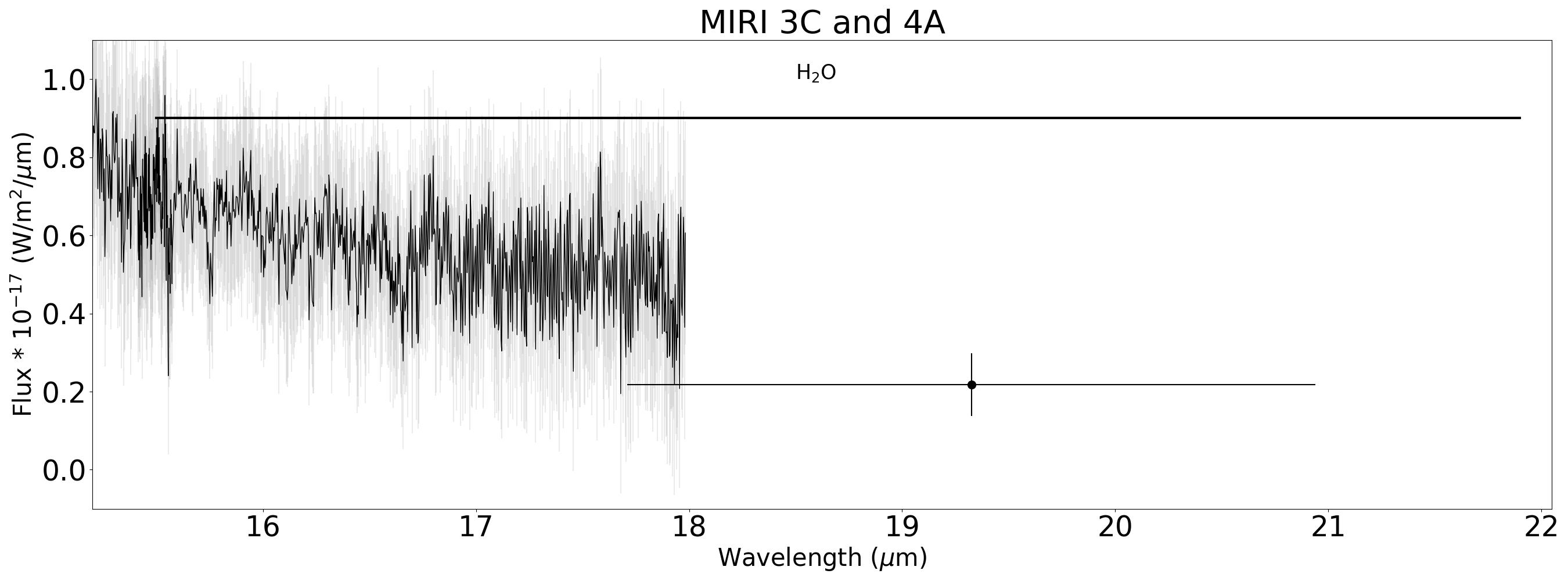}
    \caption{\textit{JWST}/MIRI spectrum of  VHS 1256 b, with important molecular gases highlighted. Molecules were identified via visual comparison with template spectra.}
    \label{fig:MIRI Features}
\end{figure*}

This spectrum is one of the highest signal-to-noise and broadest spectral coverage dataset of a brown dwarf or planetary mass companion to-date. With a spectral resolution up to $\sim$3000, a wealth of atmospheric molecular features are revealed. In this section, we visually identify absorption features in the spectra and in some cases compare them to molecular cross sections. In each subsection we discuss the molecules found and how they compare to similar brown dwarfs. The full VHS 1256 b spectrum displays several absorption features from atmospheric gases that have been previously observed in brown dwarf and extended mid-infrared coverage that can constrain methane, water, and carbon monoxide. There is evidence for carbon dioxide as well. Across the spectrum there are at least six detection regions of water, one visible methane feature, and two detections of carbon monoxide, which probe different pressure levels along the object's pressure-temperature profile. From 8 $\mu$m to 12 $\mu$m the spectral slope shows evidence of a silicate cloud. These atomic, molecular, and cloud features are highlighted in Figures~\ref{fig:NIRSpec Features} and \ref{fig:MIRI Features}.

\subsection{Near-Infrared Alkali Lines}
\begin{figure*}
    \centering
    \includegraphics[width = 5 in]{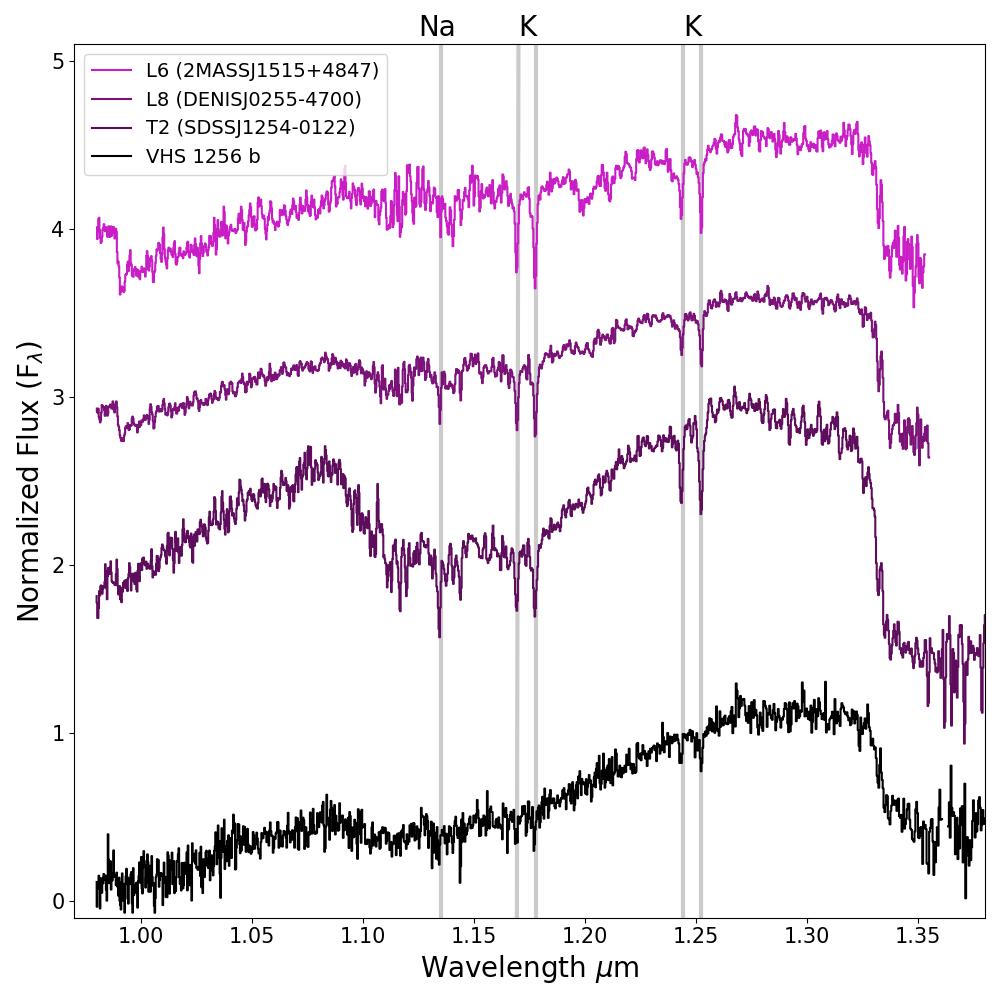}
    \caption{Template brown dwarf spectra from \cite{2013ApJ...772...79A} (colored) compared to VHS 1256 b (black). Field brown dwarfs that possess high surface gravities display broader Na and K lines than VHS 1256 b, implying the low surface gravity of VHS 1256 b.}
    
    \label{fig:gravity lines}
\end{figure*}

Late M dwarf stars and L spectral type brown dwarfs display absorption features from neutral gases such as vanadium oxide (VO), iron hydride (FeH), potassium (K), Titanium Oxide (TiO), and sodium (Na).  These absorption features can be used to infer surface gravity. Broader and deeper absorption features from these molecules and atoms correspond to higher surface gravity and older ages \citep{2013ApJ...772...79A, 2004ApJ...600.1020M}. The 1~$\mu$m to 1.35~$\mu$m portion of the \textit{JWST}/NIRSpec spectrum (Top: Figure \ref{fig:NIRSpec Features}) holds two K doublets and a Na line that indicate a relatively low surface gravity, which is consistent with VHS 1256 b's placement on a color-magnitude diagram. Figure~\ref{fig:gravity lines} shows a comparison of VHS 1256 b's Na and K lines with those of field brown dwarfs.  Both absorption lines appear narrower than the same lines in a similar spectral type field brown dwarf, 
indicating a low surface gravity for VHS 1256 b. The first published near-infrared spectrum of VHS 1256 b in \cite{2015ApJ...804...96G} showed no K doublet features and no detection of the 1.134 $\mu$m Na line, possibly due to insufficient resolution. Follow-up work by \cite{2022arXiv220706622P} obtained medium resolution (R$\sim$8,000) spectra of VHS 1256 b, at a resolution higher than the \textit{JWST} spectrum presented here. The Na line at 1.134 $\mu$m  and K doublets at 1.173 $\mu$m and 1.248 $\mu$m are easier to distinguish from the continuum in the \textit{JWST} spectrum, however the published spectrum from \cite{2022arXiv220706622P} likely has the resolution to capture a broader range of absorption features such as Fe, FeH, and TiO. However, the standard NIRSpec pipeline procedures are still in development. Once these parameters have been refined, future reductions of the \textit{JWST} spectrum may lead to the detection of finer features with higher signal-to-noise ratios.

\subsection{Water}
All spectral types of brown dwarfs possess absorption features due to water vapor in the near and mid-infrared. The shape and depth of these features are primarily determined by the effective temperature or spectral type of a brown dwarf \citep{2013ApJ...772...79A}. Previously published space-based spectroscopic infrared observations of VHS 1256 b and other brown dwarfs are sensitive to water but often limited to resolutions of a few hundred \citep{2020AJ....160...77Z}. The sensitivity and added resolution of the \textit{JWST} data provide crucial details regarding the presence of water vapor in the atmospheres of brown dwarfs. Water absorption bands are present in the VHS 1256 b spectrum at 1.3 $\mu$m - 1.6 $\mu$m, 1.7 $\mu$m - 2.1 $\mu$m, and shape the spectrum beyond 10 $\mu$m. Water overlaps with carbon monoxide (CO) between 2.2 $\mu$m and 2.6 $\mu$m and between 4.3 $\mu$m and 5 $\mu$m. All water features are labeled in Figures~\ref{fig:NIRSpec Features} and \ref{fig:MIRI Features}. 

\subsection{Methane}
Brown dwarfs below an effective temperature of $\sim$1400~K begin to display methane absorption features, as methane is favored in the chemical reaction between methane and carbon monoxide at these temperatures \citep{2002Icar..155..393L}. Methane absorption at 1.67~$\mu$m is also a typical T-dwarf signature \citep{2005ApJ...623.1115C}.  

We have an absorption feature in the spectrum at 1.66~$\mu$m in VHS 1256 b, however this feature is slightly blue-ward of the typical 1.67~$\mu$m location, but also too broad to be FeH absorption that is seen in warmer L-dwarfs \citep{2005ApJ...623.1115C}. This feature also coincides with excess amplitudes from oscillations in the extracted dithers at wavelengths between 1.65~$\mu$m and 1.75~$\mu$m. Only dithers 2 and 3 of the total four show no oscillations in the extracted spectra, but the slight depression remains when only using dither 2 and 3 of the observations. The crest-to-crest width of the oscillation waves in the calibrator star between 1.65 $\mu$m and 1.75$\mu$m is $\sim$.01$\mu$m - .02$\mu$m, and the width of the potential methane feature is .02$\mu$m. Previously published near-infrared spectra show no methane absorption at 1.67~$\mu$m \citep{2022arXiv220706622P, 2015ApJ...804...96G}. If this feature remains after further pipeline updates the opacity source causing this feature will need to be validated with further atmospheric modeling work.

We also find a relatively shallow methane feature between 2.8 $\mu$m and 3.8 $\mu$m that is consistent with the previously published L-band spectrum in \cite{2018ApJ...869...18M}. Methane absorption at $\sim$3.3 $\mu$m is more prominent in brown dwarfs than methane absorption at 1.65 $\mu$m, and is seen even in mid-L-dwarfs \citep{2000ApJ...541L..75N}. In this effective temperature range, methane has a significant opacity contribution between 7 $\mu$m and 9 $\mu$m, but its overall strength compared to water and other molecules is smaller. This region does not visually mirror the predicted opacity profile of methane for this reason. 

All of the near-infrared and mid-infrared methane features appear depleted relative to similar temperature brown dwarfs, indicating that disequilibrium chemistry is influencing the apparent abundance of methane in the upper atmosphere. We will discuss the degree of atmospheric mixing required to recreate these methane features with models in Section~\ref{sec:modeling}.

\subsection{Carbon Monoxide}
Carbon monoxide (CO) is a common near-infrared spectral feature in very low mass stars and brown dwarfs with effective temperatures above $\sim$1400 K \citep{2002Icar..155..393L, 2005ApJ...623.1115C}. Carbon monoxide produces a strong feature centered at 4.6 $\mu$m in warm brown dwarfs. Cooler brown dwarfs that have disequilibrium chemistry driven by atmospheric mixing \citep{2012ApJ...760..151S} also display this feature. VHS 1256 b's effective temperature is cool enough where a small degree of carbon monoxide will be detected at 2.3 $\mu$m. The depth and shape of the central feature is similar to previously published K-band spectra in \cite{2022arXiv220703819H} and \cite{2022arXiv220706622P}. We also detect a densely packed set of carbon monoxide features around 4.6 $\mu$m in VHS 1256 b's spectrum. It is the best resolved, highest signal-to-noise, set of features from the fundamental carbon monoxide bandhead compared to previously published space-based brown dwarfs spectra and photometry from AKARI \citep{2012ApJ...760..151S} and Spitzer \citep{2006ApJ...651..502P}.

\subsection{Carbon Dioxide}
\begin{figure*}
    \centering
    \includegraphics[width = 5 in]{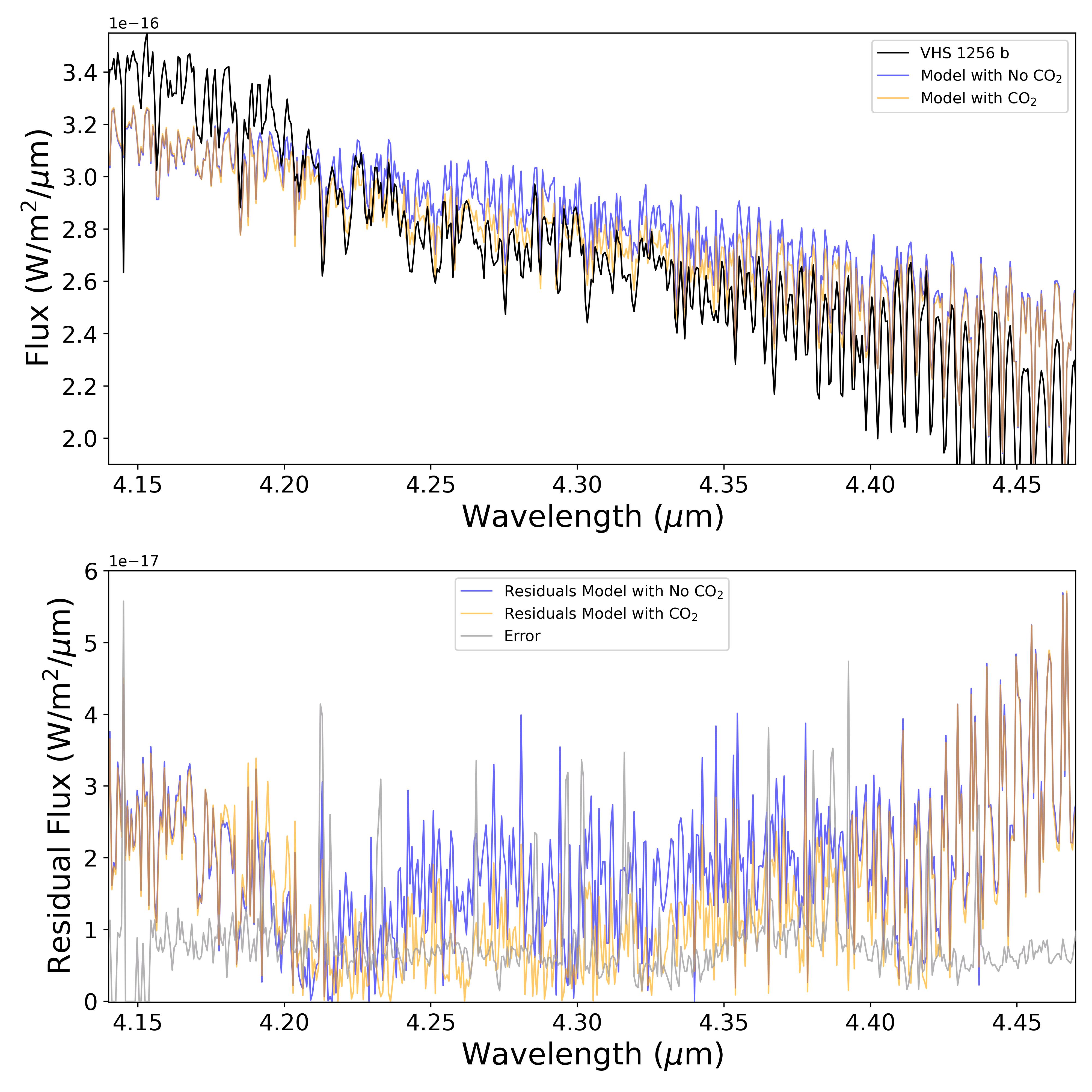}
    \caption{The \textit{JWST}/NIRSpec spectrum of VHS 1256 b compared to two atmospheric models that contain CO$_{2}$ (orange) and a model with no opacity from CO$_{2}$ (blue). The data show excess absorption at 4.2 $\mu$m that shows evidence of CO$_{2}$ absorption. The bottom panel shows residuals between the data and the models with the error plotted in gray. No CO$_{2}$ shows a larger residual than a model including the opacity of the molecule.}
    
    \label{fig:CO2}
\end{figure*}

Carbon dioxide (CO$_{2}$) is another carbon- and oxygen-bearing gas  that can provide insight into metallicity or atmospheric effective temperature \citep{2011ApJ...734...73T,2012ApJ...760..151S, 2002Icar..155..393L}. The CO$_{2}$ feature at 4.2 $\mu$m is inaccessible from ground-based observatories and has only been detected in a handful of mid-infrared observations of T-type brown dwarfs \citep{2011ApJ...734...73T,2012ApJ...760..151S}. We show evidence for this CO$_{2}$ feature in the \textit{JWST} spectrum of VHS 1256 b by comparing two versions of an atmospheric model, one with CO$_{2}$ opacities added and one without, as shown in Figure~\ref{fig:CO2}. The model comparisons reveal a slight absorption feature where CO$_{2}$ influences the spectrum from 4.2 $\mu$m to 4.4 $\mu$m. In M-dwarfs and relatively warm L-dwarfs, CO$_{2}$ is theorized to primarily trend with effective temperature, while at lower effective temperatures where methane should be the dominant carbon-bearing gas CO$_{2}$ will vary with both atmospheric pressure and temperature \citep{2002Icar..155..393L}. The presence of CO$_{2}$ in VHS1256 b's spectrum would not be not surprising, but needs to be verified with more detailed atmospheric analysis.

\subsection{Clouds}
The transition between red to blue near-infrared colors at the L-to-T transition shown in Figure~\ref{fig:CMD} is likely driven by the condensation and eventual descent below the photosphere of silicate grains composed of enstatite, forsterite, or quartz as these objects cool. Low surface gravity brown dwarfs and directly imaged exoplanets can retain their silicate clouds at lower effective temperatures, producing their redder colors compared to field counterparts. VHS 1256 b displays a significant silicate cloud feature in the \textit{JWST}/MIRI spectrum from 8 $\mu$m to 11 $\mu$m when compared to a standard brown dwarf and the relatively red brown dwarf 2MASSW J2224438-015852 (2M2224-0158) discovered in \cite{2000AJ....120..447K}, that is an outlier along the L-to-T transition (Figure~\ref{fig:silicate versus Spitzer}). Compared to 2M2224-0158, VHS 1256 b shares the same spectral shape across 8 $\mu$m to 11 $\mu$m indicating an absorption feature due to silicate clouds of a similar composition. The best fit cloud model for 2MASS 2224-0158 from \cite{2021MNRAS.506.1944B} was a combination of enstatite (MgSiO$_{3}$), quartz (SiO$_{2}$), and a higher pressure iron (Fe) cloud. 

\subsection{Other Molecular Gases and Clouds}
VHS 1256 b's spectrum shows evidence of disequilibrium chemistry based on methane and carbon monoxide, therefore we explored the potential presence
of other disequilibrium molecules. 

From the equilibrium Sonora Bobcat models \citep{2021ApJ...920...85M}, we scaled different molecular abundances to search for species that could appear in dis-equilibrium conditions. In VHS 1256 b's spectrum, there are no obvious signs of acetylene (C$_{2}$H$_{2}$), ethylene (C$_{2}$H$_{4}$), ethane (C$_{2}$H$_{6}$), hydrogen sulfide (H$_{2}$S), phosphine (PH$_{3}$), or hydrogen cyanide (HCN). Outside of the features listed prior to this subsection, more detailed analysis and approaches such as retrievals and cross correlation will need to be applied to the \textit{JWST} VHS 1256 b spectra to fully understand the object's atmospheric chemistry.

\section{Atmospheric Modeling}
\label{sec:modeling}

We used the long-developed \texttt{EGP} substellar code \citep{Marley96,Marley_2012,fortney2005comp,fortney2007planetary,fortney08,Morley14,2021ApJ...920...85M,karalidi21,Mukherjee22} to find the best-fit forward model for the observed \textit{JWST} spectrum.  This lineage of codes parametrizes the cloudiness of an object using the sedimentation parameter $f_{\rm sed}$ and parameterizes the mixing strength using the eddy diffusion coefficient K$_\mathrm{zz}$ \citep{2001ApJ...556..872A}.  A higher value of $f_{\rm sed}$ describes less optically thick clouds with larger particles: lower values produce thicker clouds with smaller particles.  The eddy diffusion coefficient K$_\mathrm{zz}$ has higher values if atmospheric mixing is strong, potentially driving the atmosphere out of chemical equilibrium \citep{Hubeny07, Saumon06} and also extending the extent of atmospheric cloud decks.  The goal of our atmospheric modeling efforts is to capture the prominent atmospheric physics and chemistry that influences the spectrum of VHS 1256 b and the \texttt{EGP} family of codes is well-suited to this task, although modeling VHS 1256 b's full spectrum required careful fine tuning of parameters to capture the object's disequilibrium chemistry and cloudiness.

We used \texttt{PICASO 3.0} \citep{Mukherjee22}, a modified Python version of the Fortran-based \texttt{EGP} code that includes the capability of self-consistently modeling both disequilibrium chemistry and clouds simultaneously with a pressure-dependent K$_\mathrm{zz}$ profile. The details of the code are described in \citet{Mukherjee22}, used opacity sources are detailed in Table 3 in \citet{Mukherjee22}.  We first explored a small grid with temperatures running from 900 K to 1600 K at intervals of 100 K and $f_{\rm sed}$ values of 1, 2, 3, and 8. Each atmospheric model has 90 layers spaced in atmospheric pressure. The spectra were post processed to mirror abundances of an atmosphere with two different uniform K$_\mathrm{zz}$ profile values, 10$^{5}$  cm$^{2}$ s$^{-1}$ and 10$^{8}$  cm$^{2}$ s$^{-1}$. The smaller value of K$_\mathrm{zz}$ corresponds to the theoretical estimated K$_\mathrm{zz}$ \citep{2014ApJ...797...41Z} value and the higher value matches previously published mixing values estimated from the 3 $\mu$m methane feature in \cite{2018ApJ...869...18M}. The best attempt at a fine-tuned model that matches the overall shape of the spectrum had parameters T$_{\rm eff}=$1100 K, $f_{\rm sed}=$1, log(g) = 4.5, and K$_{zz}=$10$^{8}$  cm$^{2}$ s$^{-1}$. However, the best fit model from this initial grid produced too much flux in the near-infrared despite matching the majority of the molecular absorption features. The estimated temperature is similar to effective temperatures (1122 $\pm$ 16K and 1171 $\pm$  17 K), derived in \cite{2022arXiv220808448D} using atmospheric evolution models from \cite{2008ApJ...689.1327S}. Our forward modeling approach produces lower effective temperatures compared to the forward model derived effective temperatures in \cite{2022arXiv220703819H} (1200 K) and \cite{2022arXiv220706622P} (1380 K) but higher than the temperature derived in \cite{2020AJ....160...77Z} (1000 K).

\begin{figure*}
    \includegraphics[width = 6.68 in]{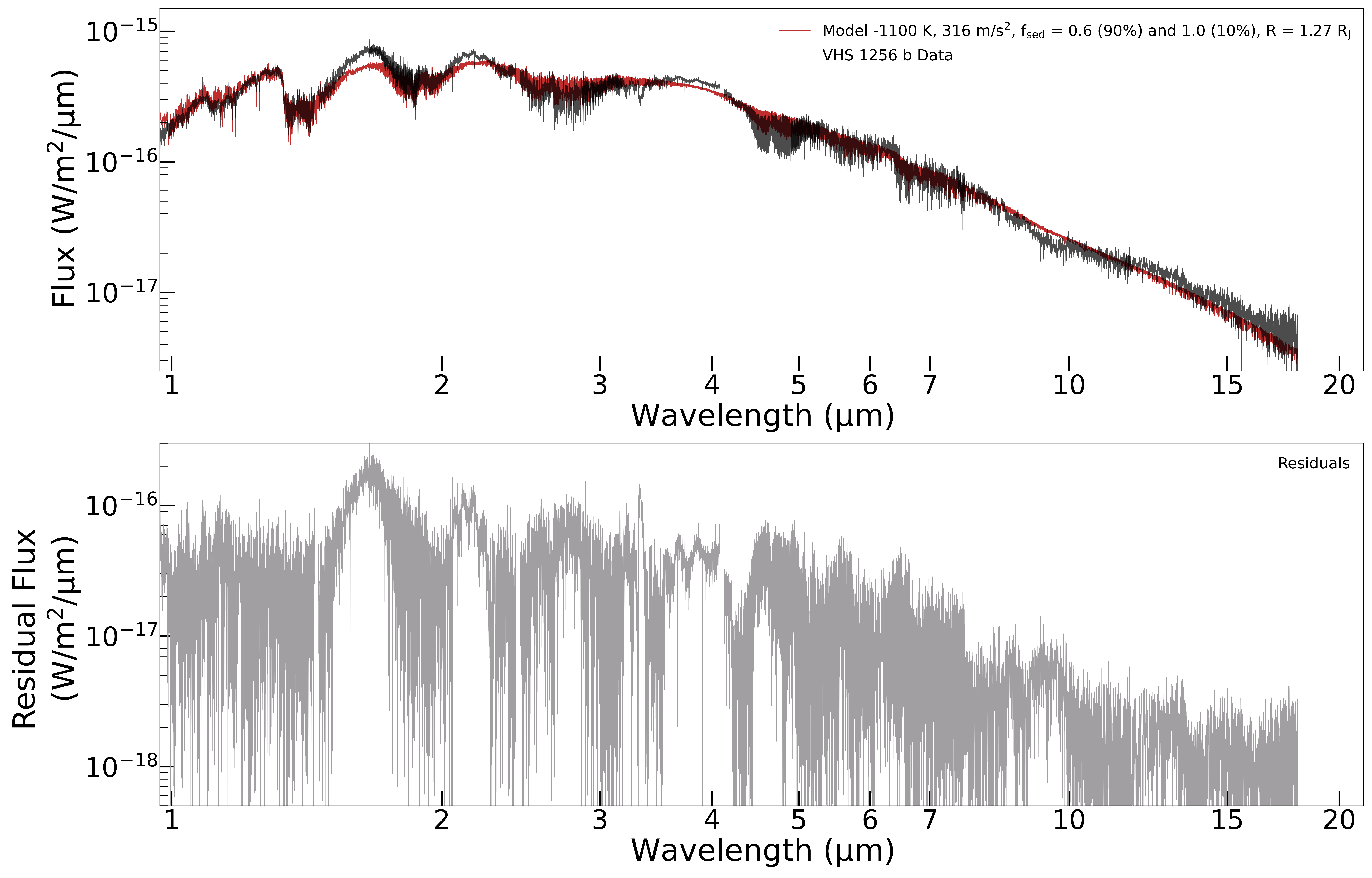}
    \caption{(Top  Panel) The best fit model atmospheric model (red) compared to the VHS 1256 b spectrum (black). The residual fluxes between the best attempt model and VHS 1256 b's spectrum. The largest discrepancies are in the near-infrared and within absorption lines that are muted due to the model's cloudiness. }
    \label{fig:best fit spectra}
\end{figure*}

\begin{figure*}
    \includegraphics[width = 6.68 in]{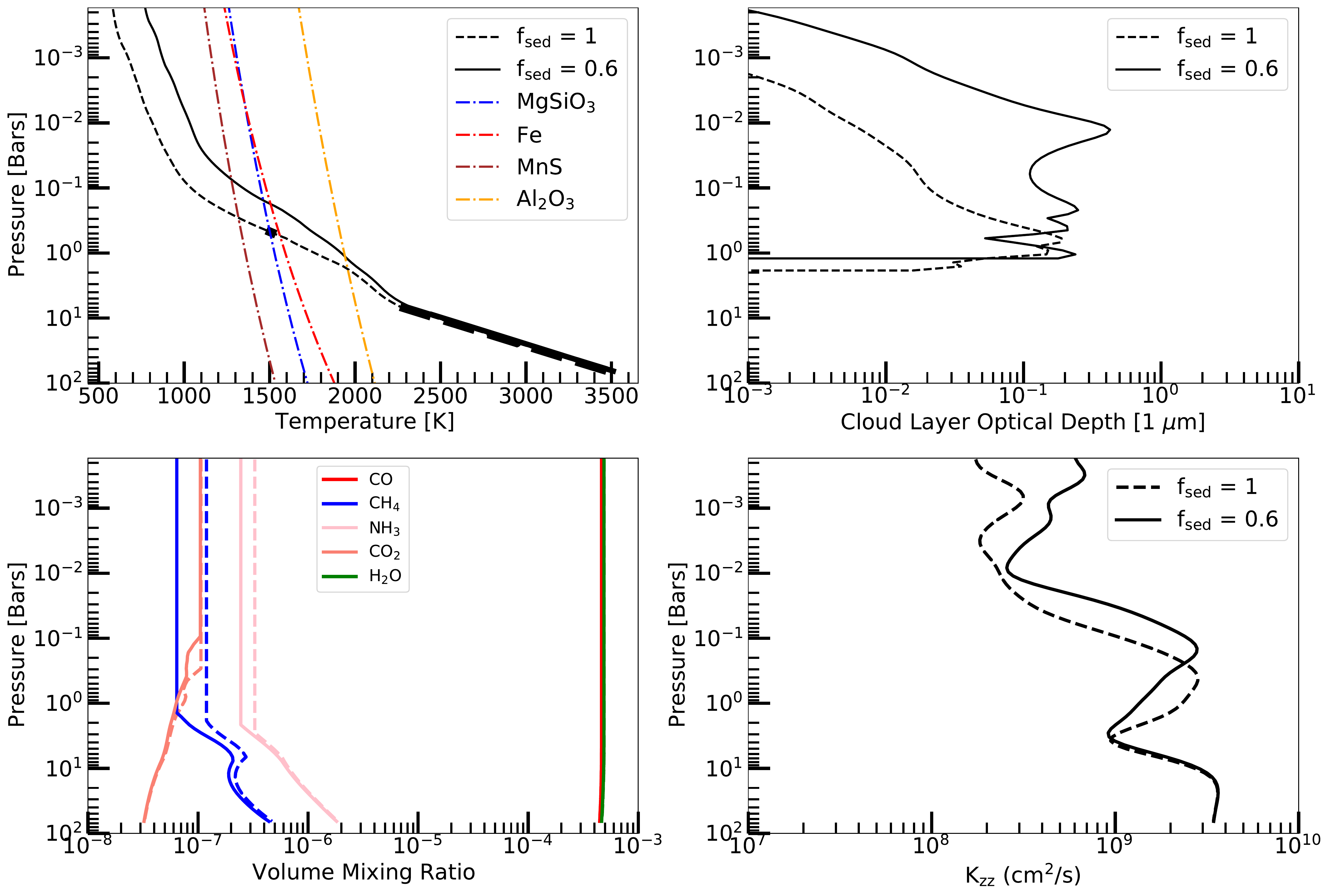}
    \caption{Atmospheric parameters of VHS 1256 b. Top left: The pressure temperature profile (black) with convective regions highlighted with thicker black lines. The condensation curves of cloud opacities included within the model are shown as colored lines. Top Right: The optical depth profile of the clouds used in the hybrid best fit model. Lower $f_{\rm sed}$ values produce a higher opacity through a larger region of the atmosphere compare to higher $f_{\rm sed}$ values.  Lower Left: The abundance of different atmospheric molecules as a function of pressure. Atmospheric mixing influences abundances primarily below log(P)= 0 bars. Lower Right: Atmospheric mixing, K$_\mathrm{zz}$, as a function of pressure. Stronger mixing occurs at higher pressures.} 
    \label{fig:best fit chemistry}
\end{figure*}

In order to better fit the observed near-infrared flux, we adopted a two--cloud model and added a self consistent treatment of atmospheric mixing to estimate K$_{zz}$ as a function of pressure. This modeling approach in \texttt{PICASO 3.0} simultaneously includes the heating/cooling due to clouds and mixing induced disequilibrium chemistry while calculating the atmospheric structure of the object. Instead of a single cloud with an $f_{\rm sed}$ of 1, we used a mixture of two clouds, with 90$\%$ of the modeled clouds having $f_{\rm sed}$ = 0.6 and 10 $\%$ of the modeled clouds having $f_{\rm sed}$ = 1.0. The clouds described by the lower $f_{\rm sed}$ = 0.6 value produce a thicker, deeper cloud layer that helps match the model to the observed spectral shape of VHS 1256b. The fluxes for each of these cloudy disequilibrium chemistry models were first calculated separately and were then linearly combined using,
\begin{equation}\label{eq:fluxcombined}
    F_{total}= fF_{1}+(1-f)F_{2}
\end{equation}
to obtain the total flux ($F_{total}$) which is shown in Figure~\ref{fig:best fit spectra}. In Equation \ref{eq:fluxcombined}, $f$ represents the fractional coverage of one of the cloudy models, $F_{1}$ and $F_{2}$ represents the fluxes calculated from models with clouds with a particular $f_{\rm sed}$. The use of two clouds is motivated by the object's measured variability, which is caused by changes in the surface brightness potentially created by moving cloud patches. Using a self consistent and pressure dependent approach to estimate atmospheric mixing, we find K$_\mathrm{zz}$ values ranging from 10$^{8}$  cm$^{2}$ s$^{-1}$ to 10$^{9}$  cm$^{2}$ s$^{-1}$ along the pressure temperature profile. The best fit spectrum using these parameters is shown in Figure~\ref{fig:best fit spectra} and the atmospheric profile and chemistry are described in Figure~\ref{fig:best fit chemistry}.

The best fit atmospheric model matches the overall shape of the 1 $\mu$m - 20 $\mu$m spectrum of VHS 1256 b with discrepancies occurring where the silicate cloud feature appears and also around the peak from 1.5 $\mu$m to 1.8 $\mu$m, where the spectrum is shaped by water and collisionally induced absorption from hydrogen. The major equilibrium and disequilibrium absorption features described in section~\ref{sec:molecular features} are not exactly matched and dampened because of atmospheric clouds. The chemical abundance profile of the best fit model demonstrates that water and carbon monoxide are relatively insensitive to atmospheric mixing.  The volume mixing ratios of methane, ammonia, and carbon dioxide are affected by atmospheric mixing, however, based on Section~\ref{sec:molecular features}, only carbon dioxide and methane produce visible spectral features.

\section{Discussion} \label{sec:discussion}
\subsection{Bolometric Luminosity and Mass} \label{sec:Lbol and Mass}

Integrating over VHS 1256 b's \textit{JWST} spectrum and using the best fit model described in the previous section to estimate flux from wavelength ranges not covered by our \textit{JWST} spectrum, we find a bolometric luminosity of $log\left(\frac{L_{bol}}{L_{\odot}}\right)$ = -4.55$\pm$0.009, adopting the \textit{Gaia} Early Data Release 3 distance measurement to the system of 21.15$\pm$0.22 pc \citep{2021A&A...649A...1G}.  Our \textit{JWST} spectrum covers most of the full luminous range of VHS 1256b; 98$\%$ of the measured luminosity is derived directly from the spectrum, with only 2$\%$ extrapolated from model fits to wavelengths outside of our \textit{JWST} wavelength coverage. So much of the SED is covered by measurements and so many independent bands are included that the statistical/random error on our bolometric luminosity value is very small. With a precise distance measurement from \citet{2021A&A...649A...1G} for this system, the most significant uncertainty will stem from the absolute flux calibration of the spectrum using the A3V star TYC 4433-1800-1 (see Section~\ref{sec:reduction}).  Over the course of the \textit{JWST}'s operation a dedicated absolute flux calibration program is planned to enable better than 2$\%$ flux calibration over a wide range of objects \citep{2022AJ....163..267G}.  Thus, to set the error on our bolometric luminosity measurement at the current early stage of the \textit{JWST} mission, we estimate a conservative error on the absolute flux calibration of 3$\%$ and add this in quadrature with the error on the integrated spectrum.

Adopting the \textit{Gaia} DR3 distance for the system, a spectral type of L7$\pm$1.5 and the VISTA Hemisphere survey $K_S$ photometry from \citet{2015ApJ...804...96G}, and a bolometric correction $BC_{Ks}$ based on the polynomial relationship for young M to T type objects from \citet{2015ApJ...810..158F}, we find a photometric bolometric luminosity of $log\left(\frac{L_{bol}}{L_{\odot}}\right)$ = -4.60$\pm$0.05, in good agreement with the value derived from our \textit{JWST} spectroscopy.  However, recent bolometric luminosity estimates from ground-based, near-IR spectroscopy for VHS 1256 b are slightly fainter than the value we find with \textit{JWST}: \citet{2022arXiv220703819H} find a best value of  $log\left(\frac{L_{bol}}{L_{\odot}}\right)$ = -4.67, within a range from -4.6 to -4.7, while \citet{2022arXiv220706622P} find $log\left(\frac{L_{bol}}{L_{\odot}}\right)$ = -4.67$\pm$0.07.  The bolometric luminosity estimate from \citet{2022arXiv220706622P} adopts a distance of 22.2$^{+1.1}_{-1.2}$ pc from \citep{2020RNAAS...4...54D}, rather than the more accurate \textit{Gaia} eDR3 measurement. Adjusting to the \textit{Gaia} distance measurement leads to a slightly smaller bolometric luminosity, further increasing the tension between the \textit{JWST} measurement and ground-based spectroscopic measurements.

\begin{figure}
    \centering
    \includegraphics[width=\columnwidth]{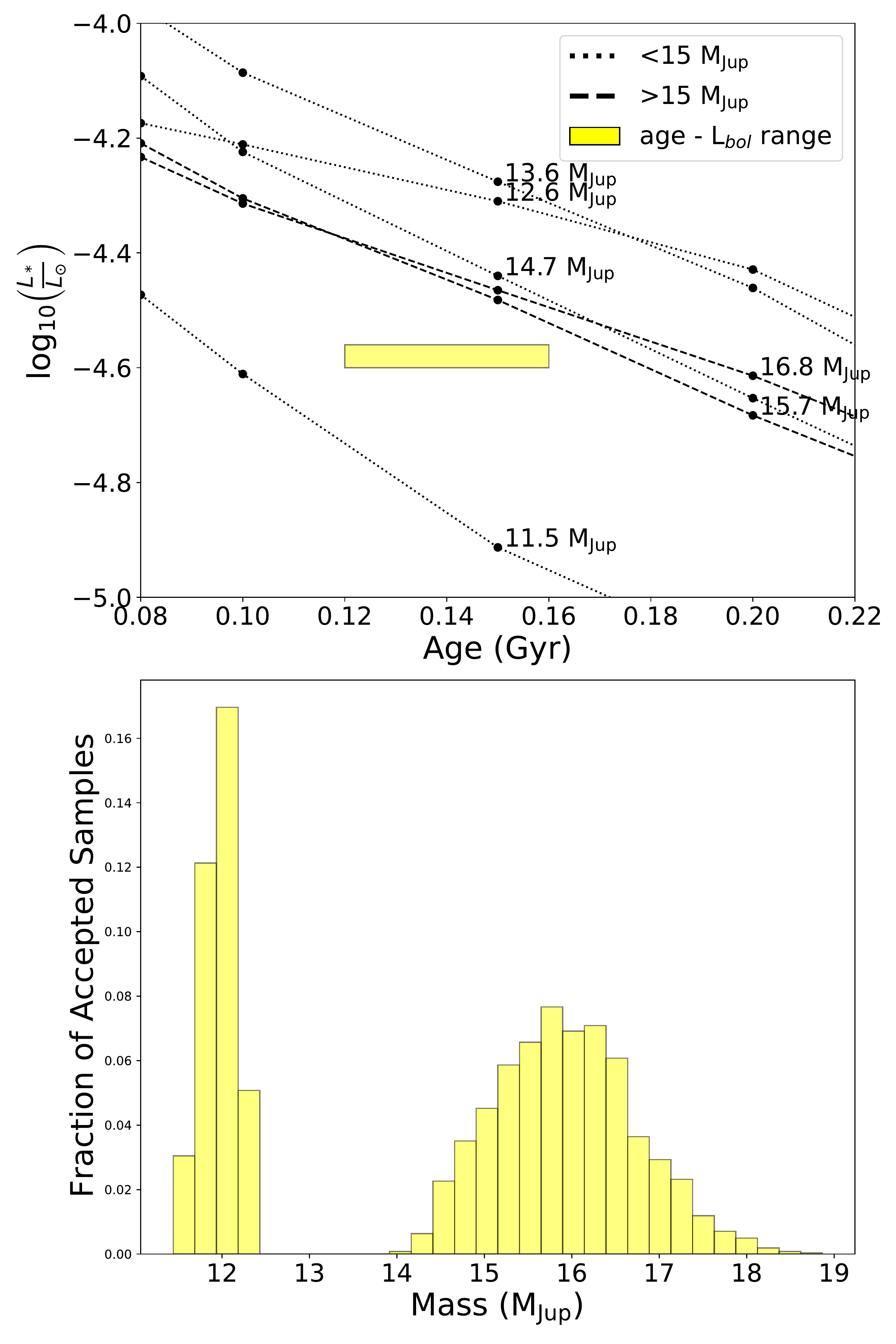}
    \caption{\textit{Top:} Mass vs. luminosity tracks from the hybrid cloud grid from \citet{2008ApJ...689.1327S}.  Models for masses $<$15 M$_\mathrm{Jup}$ are shown as dotted lines; models for masses $>$15 M$_\mathrm{Jup}$ are shown as dashed lines.  The expected age-bolometric luminosity range of VHS 1256b is shown by the yellow rectangle.  At the ages considered here, lower mass objects burning deuterium may have higher luminosities than higher mass objects which have already completed deuterium burning.   
    \textit{Bottom:}
    Histograms of the final sets of accepted masses drawn from the same grid.  The distribution is bimodal, with peaks both above and below the deuterium mass burning limit.  While we cannot conclusively determine whether VHS 1256b lies above or below the deuterium mass burning limit, all accepted samples had masses $<$20 M$_{Jup}$.}
    \label{fig:massranges}
\end{figure}

\citet{2022arXiv220808448D} determine an updated age for the system of 140$\pm$20 Myr from dynamical masses derived from orbital fits to the inner AB binary. VHS 1256 b sits close to the deuterium burning limit -- if VHS 1256b is slightly above the deuterium burning limit, it will be actively burning deuterium at this age. For the short duration of its deuterium burning phase, a lower mass object may be more luminous than a higher mass object at the same age, see Figure~\ref{fig:massranges}. As a given luminosity can correspond to a range of masses in this case, direct interpolation from this model grid can produce erroneous mass estimates.

To robustly estimate the mass of VHS 1256 b, we implemented a rejection sampling method similar to that described in \citet{2022arXiv220808448D}. First, we draw 1$\times$10$^6$ samples of age and mass from a Gaussian distribution in age around 140 Myr, with $\sigma$=20 Myr, and a uniform distribution in mass from 1 to 50 M$_{Jup}$.  For each age, mass sample we then interpolate a model luminosity from an evolutionary model grid, and calculate $\chi^{2}$ as:  $\chi^{2} = \frac{(L_{\rm bol,model} - L_{\rm bol,measured})^2}{\sigma_{\rm Lbol,measured}^2}$, then convert to a probability ($P$) by normalizing by the minimum $\chi^{2}$~($P = e^{-\frac{\chi^{2} - \chi_{min}^{2}}{2}}$) value among our 1$\times$10$^6$ samples.  For each sample, we also draw a uniformly distributed number from 0 to 1.  We retain the samples where the sample probability is greater than the uniformly distributed variate drawn for that sample.  

As VHS 1256b has strong evidence for the presence of clouds in its spectrum from the detection of the silicate feature at 10 $\mu$m, we implemented this rejection sampling procedure using the hybrid cloud grid of \citet{2008ApJ...689.1327S}, which includes clouds for objects with L-type spectra and clear atmospheres for objects with T-type spectra.  

A histogram of the final set of accepted masses are shown in Figure~\ref{fig:massranges}.  We find a bimodal distribution of masses, with accepted samples both above and below the deuterium mass burning limit, as also seen in \citet{2022arXiv220808448D}. The percentage of samples which fall into each of the two peaks depends strongly on the both the bolometric luminosity value and the uncertainty on that value.  While we cannot conclusively determine whether VHS 1256b falls above or below the deuterium mass burning limit, all accepted samples for this model have masses $<$20 M$_{Jup}$.

\subsection{Variability} \label{sec:Variability}

Young, planetary mass objects with mid-to-late-L spectral types are highly variable in the near-IR \citep{2015ApJ...813L..23B, 2018AJ....155...95B, 2016ApJ...829L..32L, 2020ApJ...893L..30B,2020AJ....160...77Z}, with variability amplitudes $>$5$\%$ over observations a few hours in length.  VHS 1256 b is the most variable of this cohort. Over a contiguous six-orbit observation with HST/WFC3 obtained in 2018, 
\citet{2020ApJ...893L..30B} found that VHS 1256 b varied by $>$20$\%$ between 1.1 to 1.7 $\mu$m over 8 hours.  In the near-IR, \citet{2020ApJ...893L..30B} fit the observed HST trend with a single sinusoidal model, however, in an additional 15-orbit / 42 hour HST / WFC3 observation in 2020 presented by Zhou et al. submitted, the variability observed is more complex, requiring a 3-sinusoid + slope model for a full fit.  
In the mid-IR,  \citet{2020AJ....160...77Z} obtained a contiguous 36-hour Spitzer 4.5 $\mu$m lightcurve of VHS 1256 b and found a best fit model for this lightcurve of a single sinusoid with a period of 22.04$\pm$0.05 hours (interpreted as the rotation period of the object) and a peak-to-peak amplitude of 5.76$\pm$0.04$\%$, a significantly lower variability amplitude when compared to the near-IR variability.      
Both the near-IR and mid-IR variability are likely driven by patchy thin and thick silicate clouds \citep{2013ApJ...768..121A}.  Variability is an important probe of the inhomogeneity of the top-of-atmosphere structure of these objects and it is critical to understand the intrinsic variability of this target in order to interpret our \textit{JWST} observations. To do so, we have conducted a multi-telescope, multi-epoch photometric variability monitoring campaign starting in February 2022 and continuing through July, directly covering the \textit{JWST} observation epoch, which will be presented in Biller et al. in prep.  

However, over the $\sim$4 hour timescale of our ERS observations, we expect a relatively low level of variability, based on estimates from the earlier HST and Spitzer observations.  To estimate the range of potential variability we expect during the NIRSpec observation, we drew 10000 sample 2-hour observations from the 3-sinusoid model from \cite{2022arXiv221002464Z}, and estimated the intrinsic variability occurring during each simulated observation as the maximum minus the minimum value from the model during that time span.  We used a very fine time sampling and did not realistically simulate noise or the actual cadence of our \textit{JWST} observations. Thus, this method constrains the intrinsic variability and the actual measurement of variability in any given observation would yield a smaller value than these predictions. From our simulated observations, 50$\%$ of samples varied by less than 1.5$\%$, 75$\%$ of samples varied by less than 2.5$\%$, and 95$\%$ of samples varied less than 3.7$\%$.  
For wavelengths $>$3 $\mu$m, drawing 10000 sample 2-hour observations from the single sinusoid model used to fit the Spitzer 4.5 $\mu$m lightcurve from \citet{2020AJ....160...77Z}, 50$\%$ of samples varied by less than 1.1$\%$, 75$\%$ of samples varied by less than 1.5$\%$, and 100$\%$ of samples varied by less than 1.6$\%$.  More conservatively, we estimate a maximum potential variability measurement of 5$\%$ over 2 hours for the \textit{JWST}/NIRSpec 1-3 $\mu$m observations, based on the highest amplitude variability epoch, where VHS 1256 b displayed 20$\%$ variability over 8 hours \citep{2020ApJ...893L..30B}.  Assuming the same scaling between near-IR and mid-IR lightcurves as found between the HST and Spitzer lightcurves of \citet{2020ApJ...893L..30B} and \citet{2020AJ....160...77Z}, we estimate a maximum potential variability measurement of 1.5$\%$ over 2 hours at wavelengths $>$3 $\mu$m.  Variability has not been measured beyond 5 $\mu$m for any planetary mass object, but assuming a continuing trend of decreasing variability amplitude with increasing wavelength, we expect variability at wavelengths $>$5 $\mu$m to be negligible.

\subsection{Atmospheric Chemistry} 
\label{sec:DisEq Chem}
The \textit{JWST} spectrum has molecular features that are consistent with previous findings of VHS 1256 b and other young, red late L-dwarfs having atmospheres that are out of chemical equilibrium \citep{2004A&A...425L..29C,2015ApJ...804...96G,2016ApJ...833...96L, 2018ApJ...869...18M}. The presence of CO absorption and depleted CH$_{4}$ absorption compared to equilibrium atmospheric models, supports atmospheric mixing forcing the atmosphere into chemical disequilibrium. The CH$_{4}$ absorption feature at 3.3~$\mu$m is the most prominent absorption feature stemming from disequilibrium chemistry in VHS 1256 b's spectrum and potentially other similar temperature directly imaged exoplanets. The fine tuned forward model shown in Figure~\ref{fig:best fit spectra} has an associated CH$_{4}$ abundance profile (Figure~\ref{fig:best fit chemistry}), but the model does not match the 3.3~$\mu$m CH$_{4}$ feature or many other molecular features along the spectrum, despite matching the overall SED shape. For sedimentation parameter values of $f_{\rm sed}$ = 4, CH$_{4}$ features at 1.6 $\mu$m and 7 $\mu$m appear in model spectra, but for our best fit model with $f_{\rm sed}$ $<$ 1, these features are quite muted. The estimated K$_\mathrm{zz}$ range of VHS 1256~b spans 10$^{8}$  cm$^{2}$ s$^{-1}$ - 10$^{9}$  cm$^{2}$ s$^{-1}$, which is consistent with the estimated K$_\mathrm{zz}$ of 10$^{8}$  cm$^{2}$ s$^{-1}$ from \cite{2018ApJ...869...18M}. However, not matching crucial disequilibrium absorption features is significant and means future modeling work on clouds and mixing will need to be done to estimate a meaningful K$_\mathrm{zz}$ for VHS 1256 b.

We show evidence of CO$_{2}$ being useful for reproducing the shape of the \textit{JWST} spectrum at 4.2 $\mu$m as shown in Figure~\ref{fig:CO2} and ~\ref{fig:best fit chemistry}. The portions of the \textit{JWST} spectrum surrounding the CO$_{2}$ feature are discrepant from the best fit atmospheric model. CO$_{2}$ is also influenced by disequilibrium chemistry, but the timescale of CO$_{2}$'s conversion from CH$_{4}$ is much faster than the conversion of CO to CH$_{4}$ \citep{2014ApJ...797...41Z}, therefore CO$_{2}$ quenches at higher atmospheric pressures.  Historically, disequilibrium chemistry driven by atmospheric mixing has only been detectable from a single molecule in exoplanets and brown dwarfs: either CH$_{4}$ for warmer L-dwarfs or CO for cooler T-dwarfs. \textit{JWST} has the capability to confidently detect two different molecular gases with different quench pressures between 1 $\mu$m and 5 $\mu$m. \textit{JWST} is capable of detecting CO$_{2}$ and CH$_{4}$ or CO over the entire L, T, and potentially Y-dwarf sequence, revolutionizing our understanding of disequilibrium chemistry as traced by carbon species in the atmospheres of brown dwarfs and extrasolar planets.

\subsection{Silicate Clouds} \label{sec:Silicates}
Brown dwarfs are expected to have clouds in their photospheres as their temperatures drop below the condensation curves of various species \citep{1986ApJ...310..238L,1989ApJ...338..314L,1996A&A...305L...1T,2000ApJ...542..464C,2001ApJ...556..872A}.  There are several lines of evidence for clouds in brown dwarfs, including: 

\begin{itemize}
    \item The red colors of L-type brown dwarfs, which quickly transition to blue colors as the brown dwarfs cool through the L-to-T transition and clouds are theorized to sink below the visible atmosphere \citep{1999ApJ...519..802K,2001ApJ...556..357A,2008ApJ...689.1327S}
    \item The variability seen in brown dwarfs \citep{2009ApJ...701.1534A,2015ApJ...799..154M}, which is most prominent near the L-to-T transition and can be modeled with patchy clouds \citep{2013ApJ...768..121A,2012ApJ...750..105R,2002ApJ...571L.151B,2008ApJ...689.1327S}
    \item The match in temperature between condensation curves and kinks in brown dwarf color-magnitude diagrams \citep{1994Icar..110..117F,1999ApJ...519..793L,2012ApJ...756..172M,2015ApJ...799...37L}.
\end{itemize}

There are also alternative explanations for the previous phenomena, such as temperature-pressure profiles that have been perturbed by disequilibrium chemistry \citep{2015ApJ...804L..17T}.  

Solid state spectroscopic features can be unambiguous signatures of clouds. At VHS 1256 b's temperature, the most prominent feature is expected to be from silicate particles, which have a broad 10 $\mu$m feature that is commonly seen in the interstellar medium and in the disks of young stars \citep{1984ApJ...285...89D}.  Using the \textit{Spitzer} Infrared Spectrograph, \citet{2006ApJ...648..614C} detected a ``plateau"-shaped absorption feature from 9-11 $\mu$m in the spectra of several mid-L brown dwarfs, which they attributed to small amorphous silicate particles.  A compendium of all of the \textit{Spitzer} spectra of brown dwarfs shows that silicate features are common, but not ubiquitous, for L2-L8 brown dwarfs \citep{2022MNRAS.513.5701S}.  

\begin{figure*}
    \centering
    \includegraphics[width = 6 in]{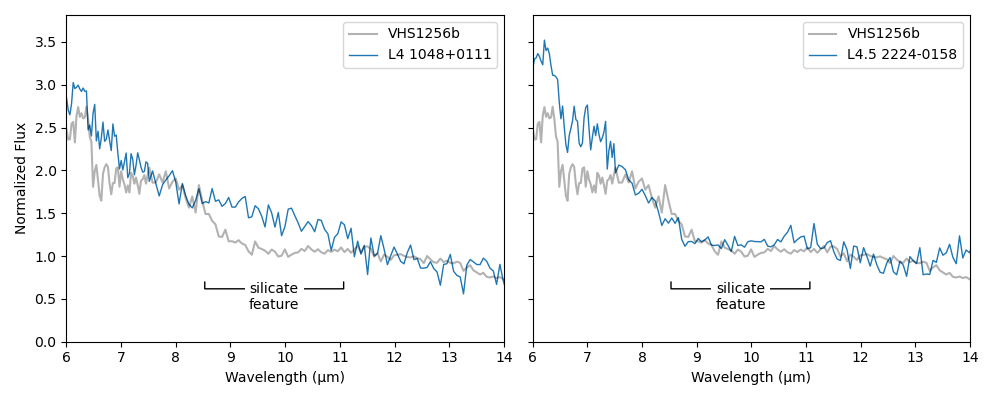}
    \caption{The \textit{JWST}/MIRI spectrum of VHS 1256 b is a good match to the \textit{Spitzer} spectrum of 2M2224-0158, which has a prominent silicate absorption feature.  2M1048+0111, which does not have a prominent silicate feature, is shown for comparison. (2M2224-0158 and 2M1048+0111 spectra from \citet{2022MNRAS.513.5701S}. }
    \label{fig:silicate versus Spitzer}
\end{figure*}

The \textit{JWST}/MIRI spectrum of VHS 1256 b shows a  prominent silicate feature compared to the \textit{Spitzer} brown dwarfs, with a shape that is well-matched to 2M2224-0158 (see Figure~\ref{fig:silicate versus Spitzer}).  \citet{2021MNRAS.506.1944B} modeled the spectrum of 2M2224-0158 and found a best-fit with clouds composed of sub-micron enstatite/silicate, quartz, and iron.  The presence of a prominent silicate feature in VHS 1256 b is strong evidence for small particles.  Detailed modeling of the cloud composition and grain size distribution will be the subject of a future paper.

Various groups have developed models for clouds in brown dwarfs and exoplanets, but they tend to focus on fitting the near-infrared (1-2 $\mu$m) part of the spectrum, where extinction from large particles (10-100 $\mu$m) mute the spectral features \citep{2000ApJ...542..464C,2001ApJ...556..357A,2008ApJ...689.1327S,2011ApJ...737...34M}.  It has long been known (e.g., from the comet literature), small particles contribute less to extinction \citep{2018ApJ...855...86G}, but they result in a much more prominent 8-12$\mu$m silicate feature \citep{2004A&A...413L..35M}. \textit{JWST}'s broad wavelength coverage allows us to constrain both populations of cloud particles.  

\section{Summary} \label{sec:Interpretation}
VHS 1256 b has several qualities that distinguish it from the typical brown dwarfs that form the L-to-T sequence shown in Figure \ref{fig:CMD}:

\begin{enumerate}
    \item VHS 1256 b is a young, low-mass object that has the characteristic red colors seen in other low-gravity objects (see Sections \ref{sec:intro} and \ref{sec:Lbol and Mass}).
    \item VHS 1256 b has the largest amplitude of variability of any substellar object to date (see Section \ref{sec:Variability}).
    \item VHS 1256 b shows disequilibrium chemistry caused by turbulent, vertical mixing (see Section \ref{sec:DisEq Chem}).
    \item VHS 1256 b has a prominent silicate feature indicating the presence of small cloud particles (see Section \ref{sec:Silicates}).
\end{enumerate}

Previous works have drawn physical connections between some of these qualities.  Young, low-mass brown dwarfs have red colors \citep{2016ApJ...833...96L,2016ApJS..225...10F}.  These same objects are more likely to have large amplitude variability \citep{2015ApJ...813L..23B,2016ApJ...829L..32L,2017ApJ...842...78V,2019MNRAS.483..480V,2020ApJ...893L..30B} and to show disequilibrium chemistry \citep{2011ApJ...733...65B,2014ApJ...797...41Z,2018ApJ...869...18M}.  Cloud models predict that turbulence and vertical mixing produce differentiated clouds with small particles at the top \citep{2001ApJ...556..872A,2006A&A...455..325H,2006A&A...451L...9H,2018ApJ...855...86G}, and objects that are reddened by extinction from large grains are more likely to have 10 $\mu$m silicate features from small grains \citep{2022MNRAS.513.5701S}.  Turbulent mixing is also likely to induce variability for objects of VHS 1256 b's temperature, which are expected to have cloudy and cloud-free regions \citep{2012ApJ...750..105R,2013ApJ...768..121A}.  Together, these properties paint a picture of a highly dynamic atmosphere, where turbulent convection drives both disequilibrium chemistry and the upwelling of condensible gasses, which form patchy silicate clouds that drive planetary variability.

\section{Conclusions}
We reduced \textit{JWST} NIRSpec IFU and MIRI MRS early release science observations of VHS 1256 b to produce the best spectrum of a planetary-mass object to date at medium resolution covering 0.97~$\mu$m - 19.8~$\mu$m. The sensitivity (NIRSpec: SNR $\sim$50 - 400, MIRI: SNR $\sim$ 7 - 20) and broad wavelength coverage of the data enabled the identification of several medium-resolution features such as water, methane, carbon monoxide, carbon dioxide, and silicate clouds within the atmosphere of VHS 1256 b. The data have sufficient signal-to-noise for forward modeling analysis to estimate the cloudiness, chemical abundance profiles, and strength of atmospheric mixing of VHS 1256 b. Our best attempt at matching the 0.97~$\mu$m - 19.8~$\mu$m spectra required combining two relatively thick cloud decks (90$\%$ $f_{\rm sed}$ = .6 and 10$\%$ $f_{\rm sed}$ = 1), low surface gravity (log(g) = 4.5), radius of 1.27 ~R$_{Jup}$, and an effective temperature of 1100 K. The derived K$_\mathrm{zz}$ profile of VHS 1256 b changes from 10$^{8}$  cm$^{2}$ s$^{-1}$ - 10$^{9}$  cm$^{2}$ s$^{-1}$, with stronger mixing generally occurring lower in the atmosphere. The best attempt model matches the overall shape of the spectrum, but does not adequately capture molecular absorption features typically used to estimate K$_\mathrm{zz}$. The luminosity of VHS 1256 b was measured to within less than a percent ($\mathrm log(L_{bol}/L_{\odot})$ = -4.55$\pm$0.009), robustly providing an upper mass limit of 20 M$_{Jup}$ for VHS 1256 b.  These initial results from the \textit{JWST} early release science observations are groundbreaking and also obtainable for numerous other nearby brown dwarfs that will be observed in future observation cycles. This observatory will be a trailblazer, pushing our understanding of atmospheric physics in planetary-companions, brown dwarfs, and exoplanets for years to come.

\section{Acknowledgements}
This project was supported by a grant from STScI (\textit{JWST}-ERS- 01386) under NASA contract NAS5-03127. This work benefited from the 2022 Exoplanet Summer Program in the Other Worlds Laboratory (OWL) at the University of California, Santa Cruz, a program funded by the Heising-Simons Foundation.  MBo acknowledges support in France from the French National Research Agency (ANR) through project grant ANR-20-CE31-0012. This project has received funding from the European Research Council (ERC) under the European Union's Horizon 2020 research and innovation programme (COBREX; grant agreement n$^{\circ}$ 885593; EPIC, grant agreement n$^{\circ}$ 819155).  This work has benefited from The UltracoolSheet at \url{http://bit.ly/UltracoolSheet}, maintained by Will Best, Trent Dupuy, Michael Liu, Rob Siverd, and Zhoujian Zhang, and developed from compilations by \citet{2012ApJS..201...19D,2013Sci...341.1492D,2016ApJ...833...96L,2018ApJS..234....1B,2021AJ....161...42B}. S-P acknowledges the support of ANID, -- Millennium Science Initiative Program -- NCN19\_171. S.M. is supported by a Royal Society University Research Fellowship. C.D. acknowledges financial support from the State Agency for Research of the Spanish MCIU through the ``Center of Excellence Severo Ochoa'' award to the Instituto de Astrofísica de Andalucía (SEV-2017-0709) and the Group project Ref. PID2019-110689RB-I00/AEI/10.13039/501100011033. B.B. acknowledges funding by the UK Science and Technology Facilities Council (STFC) grant no. ST/M001229/1.

\bibliography{submit}{}
\bibliographystyle{aasjournal}



\end{document}